\documentclass[traditabstract]{aa}
\usepackage{graphicx}
\usepackage{txfonts}
\usepackage{ulem}
\usepackage{natbib}

\usepackage{xspace}

\newcommand{\integral}{\textit{INTEGRAL}\xspace}

\newcommand{\eqpair}{\texttt{eqpair}\xspace}
\newcommand{\belm}{\texttt{belm}\xspace}

\newcommand{\lhls}{\ensuremath{l_\mathrm{h}/l_\mathrm{s}}\xspace}
\newcommand{\ls}{\ensuremath{l_\mathrm{s}}\xspace}
\newcommand{\kTbb}{\ensuremath{kT_\mathrm{bb}}\xspace}
\newcommand{\lnthlh}{\ensuremath{l_\mathrm{nth}/l_\mathrm{h}}\xspace}
\newcommand{\taup}{\ensuremath{\tau_\mathrm{p}}\xspace}
\newcommand{\Ginj}{\ensuremath{\Gamma_\mathrm{inj}}\xspace}
\newcommand{\refl}{\ensuremath{\Omega/2\pi}\xspace}
\newcommand{\cyg}{Cyg~X-1\xspace}
\newcommand{\lnth}{\ensuremath{l_\mathrm{nth}}\xspace}
\newcommand{\lBnth}{\ensuremath{l_\mathrm{B} / l_\mathrm{nth}}\xspace}
\newcommand{\kTin}{\ensuremath{kT_\mathrm{in}}\xspace}

\newcommand{\lth}{\ensuremath{l_\mathrm{th}}\xspace}

\begin{document} 

   \title{Potential origin of the state-dependent high-energy tail in the black hole microquasar Cygnus X-1 as seen with INTEGRAL}
   \subtitle{}

   \author{F. Cangemi\inst{1,2} \and
          T. Beuchert\inst{3} \and
          T. Siegert\inst{4}
          \and J. Rodriguez\inst{1} \and V. Grinberg\inst{5} \and
          R. Belmont\inst{1}  \and C. Gouiff\`es\inst{1} \and I. Kreykenbohm\inst{6} \and P. Laurent\inst{1} \and K. Pottschmidt\inst{7,8} \and J. Wilms\inst{6}\fnmsep}

\offprints{fcangemi@lpnhe.in2p3.fr ; tbeuchert@eso.org}
\authorrunning{Cangemi et al.}
\titlerunning{Origin of the Cygnus X-1 high-energy tails}

   \institute{Lab AIM, CEA/CNRS/Universit\'e Paris-Saclay, Universit\'e de Paris, F-91191 Gif-sur-Yvette, France
             \and Laboratoire de Physique Nucl\'eaire et des hautes \'energies, Sorbonne Universit\'e, 4 Place Jussieu, 75005 Paris, France\\ \email{fcangemi@lpnhe.in2p3.fr}
             \and European Southern Observatory, Karl-Schwarzschild-Straße 2, 85748 Garching bei München, Germany
             \email{tobias.beuchert@eso.org}
             \and Center for Astrophysics and Space Sciences, University of California, San Diego, 9500 Gilman Drive, La Jolla, CA 92093, USA
             \email{tsiegert@ucsd.edu}
             \and  Institut f\"ur Astronomie und Astrophysik, Universität T\"ubingen, Sand 1, 72076 T\"ubingen, Germany
             \and Dr.~Karl Remeis-Sternwarte and Erlangen Centre for Astroparticle Physics, Friedrich-Alexander Universit\"at Erlangen-N\"urnberg, Sternwartstr.~7, 96049 Bamberg, Germany
             \and CRESST, CSST, Department of Physics, University of Maryland Baltimore County, 1000 Hilltop Circle, Baltimore, MD 21250, USA
            \and NASA Goddard Space Flight Center, Astrophysics Science Division, Code 661, Greenbelt, MD 20771, USA
             }
   \date{Accepted for publication}

  \abstract{
  0.1--10\,MeV observations of the black hole microquasar Cygnus X-1 have shown the presence of a spectral feature in the form of a power law in addition to the standard black body (0.1--10\,keV) and Comptonization (10--200\,keV) components observed by \integral in several black-hole X-ray binaries. This so-called ``high-energy tail'' was recently shown to be strong in the hard spectral state of Cygnus X-1, and, in this system, has been interpreted as the high-energy part of the emission from a compact jet.}
  {This result was nevertheless obtained from a data set largely dominated by hard state observations. In the soft state, only upper limits on the presence and hence the potential parameters of a high-energy tail could be derived. Using an extended data set, we aim to obtain better constraints on the properties of this spectral component in both states.}
  {We make use of data obtained from about 15 years of observations with the \integral satellite. The data set is separated into the different states and we analyze stacked state-resolved spectra obtained from the X-ray monitors (JEM-X), the gamma-ray imager (IBIS), and the gamma-ray spectrometer (SPI) onboard.}
  {A high-energy component is detected in both states, confirming its earlier detection in the hard state and its suspected presence in the soft state with \integral, as seen in a much smaller SPI data set. We first characterize the high-energy tail components in the two states through a model-independent, phenomenological analysis. We then apply physical models based on hybrid Comptonization (\eqpair and \belm). The spectra are well modeled in all cases, with a similar goodness of the fits. While in the semi-phenomenological approach the high-energy tail has similar indices in both states, the fits with the physical models seem to indicate slightly different properties. Based on this approach, we discuss the potential origins of the high-energy components in both the soft and hard states, and favor an interpretation where \sout{the part of} the high-energy component is due to a compact jet in the hard state and hybrid Comptonization in either a magnetized or nonmagnetized corona in the soft state.}{}


   \maketitle
%

\section{Introduction}

Black-hole X-ray binaries (BHXBs) are systems with variable X-ray
emission and can transit through various spectral states during
outbursts. Because of the discovery of different types of radio relativistic jets in these systems \citep[e.g.,][]{Mirabel92, Fender99}, they are commonly dubbed \textquotedblleft microquasars\textquotedblright. The two main (canonical) states are the low-hard state (LHS) and the high-soft state \citep[HSS; see, e.g., ][for a precise definition of the states]{Remillard2006,Belloni2010}. In the LHS, the X-ray spectrum between ${\sim}1$ and ${\sim}100$\,keV is dominated by a hard power law with a photon index of $\Gamma \lesssim 1.5$ which cuts off at a few hundred keV. Compact radio jets are only observed in the LHS \citep[e.g.,][]{Fender01} and have been resolved in a couple of sources: Cygnus X-1 (Cyg X-1) and GRS 1915+105 \citep[e.g.,][]{Stirling2001, Fuchs03}. In
the HSS, the spectrum is dominated by a blackbody component peaking at ${\sim}1\,$keV. This component is associated with thermal emission from an optically thick accretion disk. Usually, no radio emission is detected in this state
\citep[e.g.,][]{Fender99,Remillard2006,Belloni2010}, with the possible exception of Cyg X-1 where tenuous radio emission was detected using very long baseline interferometry \citep{Rushton2012}. 

In several bright sources, observations made at higher energies reveal the presence of a high-energy tail above a few hundred keV that can sometimes extend to several MeV \citep{Grove1998}. In this paper we seek further insight into the state-resolved high-energy tail emission in order to discuss its physical origin which is still
subject to much debate. To do so, we focus on the most promising
candidate, \cyg, and exploit the combined capabilities of the Joint European X-Ray Monitor \citep[JEM-X,][]{Lund2003}, the  Imager on Board the \integral Satellite \citep[IBIS;][]{Lebrun2003}, and the SPectrometer onboard \integral \citep[SPI;][]{Vedrenne2003} to perform detailed spectral analysis of the broad-band $\sim$3\,keV--2\,MeV data by accumulating all existing \integral\ data of this source taken over the first 15 years of the satellite's life.

Although it was used as a prototype for the definition of the
(canonical) X-ray spectral states, \cyg is a peculiar case. Located at a distance of $1.86 \pm 0.12$\,kpc \citep{Reid2011}, it is a
persistent high-mass X-ray binary (HMXB) with a $14.8 \pm 1.0\,M_\odot$ black hole orbiting a blue supergiant star with a mass of
$19.2 \pm 1.9\,M_\odot$ \citep{Orosz2011}. The orbital plane is seen
at an inclination of ${\sim}27$\degr\ (\citealt{Orosz2011},
Miller-Jones et al., submitted), a value that we use throughout this paper. In contrast to the majority of
(transient) BHXBs, its individual states persist on much longer timescales \citep[e.g.,][]{Grinberg2013}, and even though the parameters ofx its spectral states are similar to those of other sources, Cyg X-1 does not follow the typical \textquotedblleft q-track\textquotedblright\, pattern for transients: for example, it has never returned to quiescence.

A high-energy tail above the thermal Comptonization continuum in
Cyg~X-1 was detected for the first time in both the soft and hard
accretion states by the COMPTEL instrument
\citep{McConnell2000,McConnell2002} on board the Compton Gamma-Ray Observatory (\textit{CGRO}). The tail in the LHS was later confirmed with the SPI \citep{Cadolle2006,Jourdain2012} and IBIS
\citep[][hereafter L11, R15]{Laurent2011,Rodriguez2015}, but so far
has remained undetected by \integral in the HSS \citep[][hereafter J14]{Jourdain2014}. This might be both due to the intrinsic faintness of the component and the rather short exposure time of  \integral in the HSS. Observations obtained between 2003 and 2011  contained $\sim 90$\,\% of LHS data (e.g., R15). Thanks to a spectral transition to a long-lasting HSS in 2011 \citep{Grinberg2011}, follow-up studies such as the present one will be increasingly sensitive to the high-energy tail emission of \cyg in the HSS.

Detection and investigation of the high-energy tails in both
states of \cyg with \integral\ are possible thanks the
mission's combination of a large effective area with the capability to measure polarization -- a key tool for breaking degeneracies in the quest for the physical origin of the high-energy tail emission. Several studies have confirmed a high degree of polarization for the high-energy tail in the LHS \citep[L11,][and R15 for a state-resolved polarimetric analysis]{Jourdain2012}, which is compatible with synchrotron emission from a jet \citep[observed in the source's LHS e.g.,][]{Stirling2001,Zdziarski2012}. However, this extension above the Comptonized spectrum could also originate from a nonthermal electron distribution in the corona \citep[e.g.,][]{McConnell2002,Romero2014,Jourdain2014,Chauvin2018}. In particular, \citet{DelSanto2013} separated all Cyg X-1 \integral data obtained until 2009 into 12 separate sub-states according to their hardness in the IBIS range and their 30--100\,keV spectral slope. They studied these 12 stacked spectra within the framework of hybrid models, either nonmagnetized (\eqpair) or with magnetization (\belm), and were able to obtain upper limits on the corona's magnetic field. 

We are therefore not only interested in assessing the existence of a high-energy tail in the HSS but also in investigating its origin with the extended amount of data available. Our approach, accumulating data until 2017, increases the count statistics by a significant amount over all earlier works and allows us to study the parameter space of hybrid models not only in the LHS but also, for the first time, with \integral in the HSS up to 2\,MeV. This paper is organized as follows. In Sect.~\ref{sec:obs_red}, we outline the selection criteria for data considered in this work and the extraction of data products. In Sect.~\ref{sec:analysis}, we present a set of techniques and models used to describe the data products. In Sect.~\ref{sec:discussion}, we discuss the results and in Sect.~\ref{sec:conclusion} we provide our conclusions.

\section{Observations and data reduction}
\label{sec:obs_red}

\subsection{State classification and data selection}

We consider all \cyg observations made with \integral from 2003 June 9 to 2017 November 7 (Modified Julian Day, MJD 52799 to 58054) corresponding to the satellite revolutions (hereafter revs) 54 to 1882. We restrict our analysis to ``science windows'' (hereafter scws), that is, individual \integral pointings of typically 1800--3600\,s duration, where \cyg is less than 10\degr\ off axis. This results in a total of 7288 scws corresponding to 11.4\,Ms of dead time-corrected exposure.

\begin{figure*}[h]
    \centering
    \includegraphics[width=\textwidth]{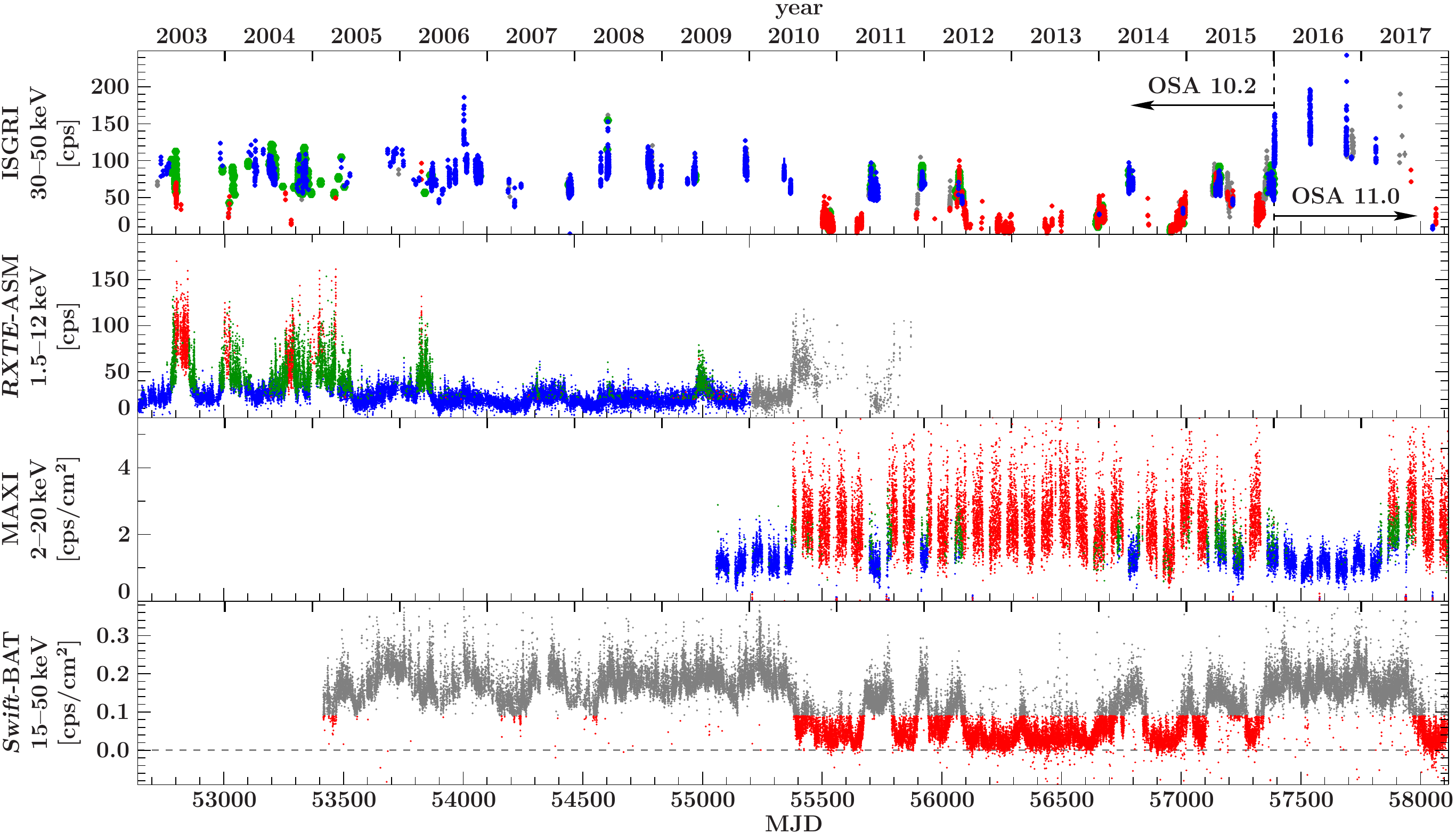}
    \caption{Spectral state evolution of \cyg between 2003 and 2017 after
      \citet{Grinberg2013}. The HSS is shown in red, the IS in green, and the LHS in blue. Unclassified data are plotted in gray.
      \textsl{Upper panel:} 30--50\,keV \textsl{INTEGRAL}/ISGRI light
      curve. The vertical dashed line indicates the change of the
      \integral\ data analysis software. \textsl{Lower three panels:}
      \textsl{RXTE}/ASM, \textsl{MAXI,} and \textsl{Swift}/BAT light
      curves of \cyg. }
    \label{fig:ibislc}
\end{figure*}

Similar to R15, we separate the whole data set into three states (soft/HSS, intermediate/IS, and hard/LHS) using the model-independent classification of \citet{Grinberg2013}. This classification defines LHS, HSS, and IS based on count rates and hardnesses measured by all-sky monitor instruments (\textit{R}XTE/ASM, \textit{MAXI}, and \textit{Swift}/BAT). Figure~\ref{fig:ibislc} shows the classification of the individual scws. This figure extends the similar Fig.~1 from \citet{Grinberg2013}, to whom we refer for the details of the methodology of the classification. As in R15, we use all-sky monitor data within 6\,hours from a given \integral scw for the classification. Such stringent time constraints on the classification are necessary because the source is known to undergo fast state transitions \citep{Boeck2011,Grinberg2013}. In total, we roughly double the total exposure time in the LHS compared to R15, and the new extended HSS covering roughly MJD 55800 to 57400 (with some excursions into the LHS, Fig. \ref{fig:ibislc}) allows us to increase the HSS exposure by a factor of three compared to R15. After MJD 57400, \integral mostly catches Cyg X-1 in the LHS (Fig. \ref{fig:ibislc}). To summarize, we obtain 3178, 2262, and 600 scws, in the LHS, HSS, and IS respectively, corresponding to 4.4\,Ms, 3.4\,Ms, and 0.9\,Ms of exposure time. A total of 1248 scws remain unclassified with this method and are not taken into account in our analysis. 

\subsection{Spectral extraction: \integral/IBIS/ISGRI}
\label{sec:isgri_data}

Data from \integral/IBIS' upper detector layer ISGRI\footnote{The \integral's Soft Gamma-Ray Imager \citep{Lebrun2003}} taken before December 2015 (\integral revolution~1627, MJD~57387) were reduced with version 10.2 of the \texttt{Off-line Scientific Analysis (OSA)} software. Data obtained after that date were extracted with \texttt{OSA\,v11}. This split is necessary because (at the time of writing) specific versions of calibration data ---specifically for the gain correction of the instrumental energy scale--- are required for both software versions\footnote{See \url{https://www.isdc.unige.ch/integral/analysis#Software} for the restrictions on the use of \texttt{OSA\,v11} with IBIS.}. The long-term IBIS light curve (with a resolution of individual scws) is shown in the top panel of Fig.~\ref{fig:ibislc}. The different gain correction used before and after December 2015 is seen as an overall increase in the 30--50\,keV count rate. For comparison, the 30--50\,keV Crab count rate is ${\sim}87\,\mathrm{counts}\,\mathrm{s}^{-1}$ for \texttt{OSA\,v10.2} and ${\sim}134\,\mathrm{counts}\,\mathrm{s}^{-1}$ for \texttt{OSA\,v11}.

IBIS/ISGRI data were reduced with the standard procedure\footnote{\label{ibis}\url{https://www.isdc.unige.ch/integral/download/osa/doc/11.1/osa_um_ibis/man.html};}. For each satellite revolution we first produce a mosaic image to identify the brightest sources (here defined as those with a detection significance greater than 7$\sigma$ in the 30--50 keV mosaic image) of the field. We then extract light curves and spectra on a scw-by-scw basis. For each scw, we construct a sky model including the  brightest sources as defined above (usually Cyg X- 3, Cyg X-2, GRO J2058+42, and KS 1947+300, in addition to Cyg X-1). Taking the background into account, the analysis software then reconstructs the sky image and corresponding source count rates by deconvolving the shadowgrams projected on the detector plane. The roughly 250 scws from the time period of the 2015 June outburst of V404~Cyg (\integral revolutions 1554 through 1558) remain unclassified and were  therefore ignored in our analysis.

Source spectra were extracted on a logarithmic grid of 60 spectral channels. In \texttt{OSA v11.0}, the spectral binning and the response matrices are calculated and provided automatically by the extraction software. In \texttt{OSA\,v10.2}, the binning is provided by the user and based on a pre-rebinned redistribution matrix file (rmf). We construct the latter to be as close as possible to the \texttt{OSA v11.0} one, resulting in spectral bin deviations of less than 0.25\,keV per channel. We then stack the individual spectra according to the version of the extraction software and following their spectral state classification. This results in a total of four stacked spectra: two per spectral state. In all spectra, we add 2\,\% systematic uncertainty to each of the channels\textsuperscript{\ref{ibis}}. For the spectral fittings, we omit the spectral channels below 25 keV for all the ISGRI spectra obtained with \texttt{OSA v10.2} and those below 30 keV for the \texttt{OSA 11.0} ones.

\subsection{Spectral extraction: \integral/SPI}
\label{sec:spi_data}

The SPI data analysis relies on the comparison of sky models and a description of the instrumental background to data in the raw format of counts per energy bin, detector, and pointing (scw). See \citet{Vedrenne2003} for details about the instruments. For the SPI extraction we include the source-intrinsic and background variability in the sky model, yielding state-resolved, long-term spectra. In the case of \cyg our sky models consist of point-sources in the $16^\circ \times 16^\circ$ field of view with time-dependent fluxes. Above ${\sim}100$\,keV, only \cyg and Cyg~X-3 are significantly detected and taken into account.

\begin{figure}
    \centering
    \includegraphics[width=\columnwidth,trim=0.00in 0.00in 0.00in 0.00in,clip=true]{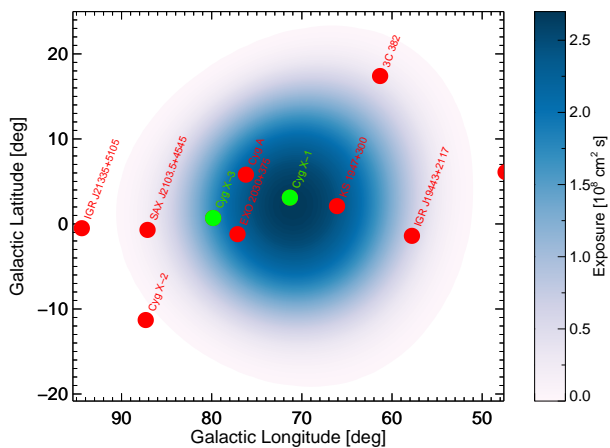}
    \caption{SPI exposure of \cyg in its soft state in units of $10^6\,\mathrm{cm^2\,s^{-1}}$. The fully coded field of view of SPI is $16^{\circ}\times16^{\circ}$. Green points indicate point sources that have to be taken into account in the SPI data analysis. Red points show other sources listed in the SPI catalog of \citet{Bouchet2011_diffuseCR} but only seen below $\lesssim$100\,keV or at specific flare times.}
    \label{fig:spiexpomap}
\end{figure}

First, we collect the good time intervals as defined for the
IBIS/ISGRI data analysis divided into soft and hard state. Because SPI is very sensitive to instrument mode changes, (solar) particle events, detector failures, and Van Allen belt passages, we apply additional filter criteria as shown in Table~\ref{tab:spi_filter}. The resulting number of pointings is 2358 and 1771 for the LHS and HSS, respectively, with a total dead-time-corrected observation time of 3.4\,Ms and 4.7\,Ms, respectively. The resulting exposure map using all good pointings is shown in Fig.~\ref{fig:spiexpomap}. This map shows \cyg in the center of the chosen exposure and which (known) sources are exposed and how they might contribute in the fit.

\begin{table}
    \caption{Selection criteria for the SPI data set based on ``science housekeeping'' (HK) parameters. Revolutions 1554 to 1558 are excluded because of the V404 Cygni outburst.}
    \label{tab:spi_filter}
    \centering
    \begin{tabular}{l l}
    \hline \hline
        HK parameter & Allowed values/ranges \\
        \hline
        SPIMode & 41 \\
        ACSRate/GeSatTot & $\left[360,400\right]$ \\
        HVDetMean & $\left[2.05,4.05\right]$ \\
        OrbitRevPhase & $\left[0.1,0.9\right]$ \\
        $\Delta$TempColdPlt & $\left[0.5,0.8\right]$ \\
        FracGoodEvtsTot & $\left[0.04,0.06\right]$ \\
        IREM TC2Rate & $<3$ \\
        IREM S32Rate & $<5$ \\
        Revolution & $\lnot\left\{ 236,354,1147,1149,1604 \right\}$\\
         & $\lnot \left\{ 1554,\ldots,1558 \right\}$ \\
        \hline
    \end{tabular}
\end{table}

We consider single events (SEs) below 530\,keV and pulse-shape discriminated (PSD) events above that energy. This removes strong ``electronic noise'' features between 1.4 and 1.7\,MeV and allows for a smooth spectral extraction\footnote{\label{spi}\url{https://www.isdc.unige.ch/integral/download/osa/doc/11.1/osa_um_spi/man.html}}. We also extract SE spectra up to 755\,keV to establish consistency between the two data types. The derived PSD fluxes are then scaled by a factor of 1/0.85 to account for the change in efficiency \textsuperscript{\ref{spi}}.

For the spectral extraction, we define 21 logarithmic energy bins between 25\,keV and 755\,keV for the SE range, and 8 logarithmic bins between 530\,keV and 2000\,keV for the PSD range. The spectra are obtained bin by bin from maximum-likelihood fits accounting for Poisson-distributed data. These fits include the two-component background model as described by \citet{Siegert2019_SPIBG}, which is based on the SPI BackGround and Response DataBase \citep[BGRDB,][]{Diehl2018_BGRDB}, as well as \cyg and Cyg~X-3. We use the \textit{OSA} \textit{spimodfit} \citep{Strong2005_gammaconti} to perform these fits. Given the large unknown in the SPI data reduction processes, and although Cyg X-1 is not formally a source sufficiently bright to present the caveats described by \citet{Roques2019} in both states, we also extract SPI spectra using different methods (see Appendix \ref{app:spi_extraction}). 

The SPI background variation timescale depends on energy and energy bin size \citep{Siegert2019_SPIBG}; additionally, the absolute fluxes of the sources may  change with time, even when they are in the same state. To determine the optimal variation timescales for both background and source at the same time, we therefore perform a grid search. To do so, we define 13 variability timescales, between one pointing (${\sim}$0.6\,h) and no variability (constant background and source), and double the times in each step (in total 169 grid points). The smaller the variability timescale, the more parameters are included in the fit. Although this results in a better likelihood, it tends to over-sample the data. In order to penalize large numbers of parameters, we search for the minimum Akaike information criterion \citep[AIC][]{Akaike1974, Cavanaugh1997} instead, which defines the fit quality of models relative to each other (see the Appendix for examples of AIC maps). From this analysis, we find that in the HSS the background variability timescale is rather constant as a function of energy, changing between 12\,h and 72\,h (${\sim}1/6$--1 INTEGRAL orbit). In the LHS, background variations are more frequent. The background variability timescale is found to decrease down to the pointing timescale below 150-200\,keV. However, at these energies the background and source patterns in addition to the large number of fitted parameters are too degenerate to determine a robust estimate for a temporal variability timescale of \cyg.

It appears that switching off a strong source variability provides similar flux normalizations to IBIS/ISGRI spectral extractions ($c_{\mathrm{ISGRI}}/c_{\mathrm{SPI}} \approx 1.0$, see Table\,\ref{table_pars}) in the LHS, but to a lesser extent in the HSS ($c_{\mathrm{ISGRI}}/c_{\mathrm{SPI}} \approx 1.3$). Above 300\,keV (400\,keV) in soft (hard) state, the number of fitted parameters for background and source are considerably smaller, also reducing the degeneracy in the maximum likelihood fits. The source spectra are then found to be steady, and the high-energy tail (Sect.~\ref{sec:phen}) is constant in time.

\begin{figure}
    \centering
    \includegraphics[width=\hsize]{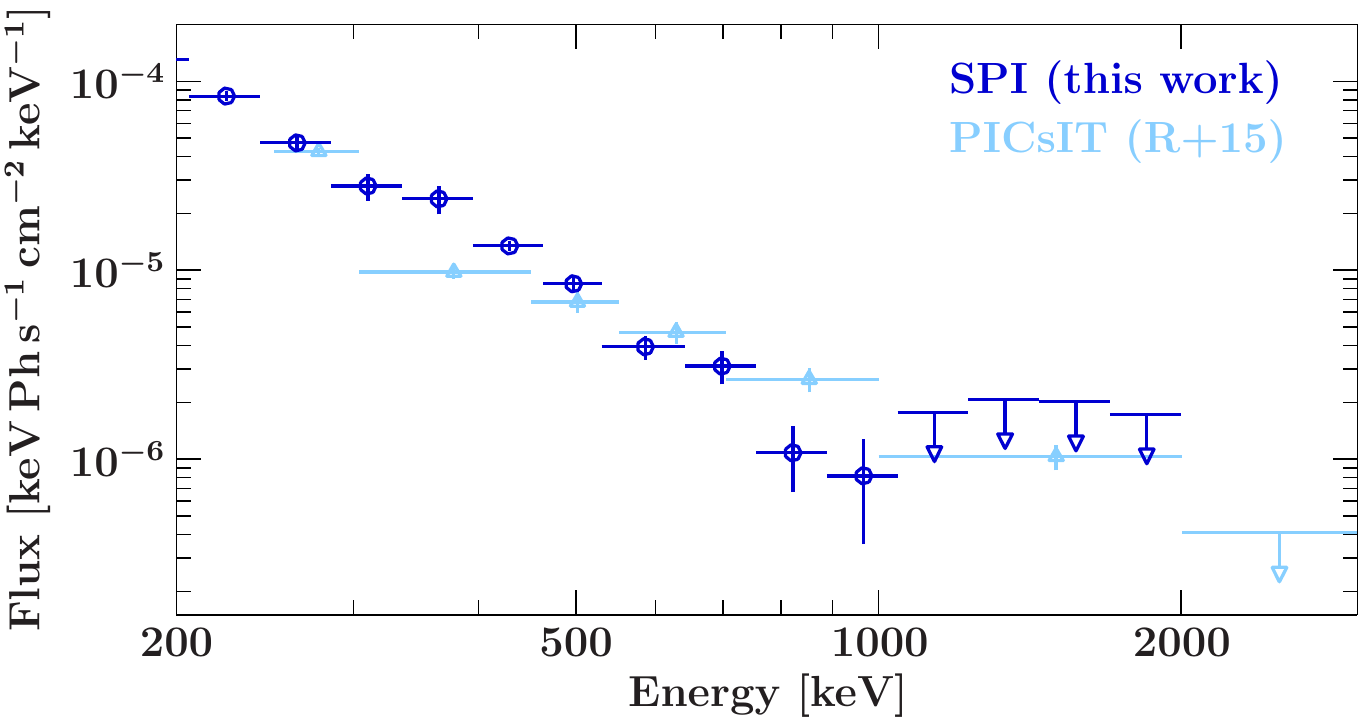}
    \caption{Comparison of our extracted hard-state \integral/SPI spectrum with \integral/IBIS/PICsIT spectra (a) extracted for the same time period as SPI and (b) for data before 2012 as published by \citet{Rodriguez2015}.}
    \label{fig:spipicsit}
\end{figure}

We compare our extracted \integral/SPI hard state spectrum with the \integral/IBIS Compton-mode spectrum published by R15 in Fig.~\ref{fig:spipicsit}. At the time, the spectrum was blamed for being very hard. Here, we observe a softer spectrum as extracted with SPI compared to the Compton-mode spectrum previously published by R15.

\subsection{Spectral extraction: \integral/JEM-X}
\label{sec:jem-x_data}

For the Joint European X-ray Monitor spectral extraction, we restrict our selection to scws where the source is in the field of view of the 
JEM-X unit 1, which is where the source is less than 5$^{\circ}$ off-axis. This selection results in 836 and 1112 scws for the LHS and 
HSS, respectively. The data from JEM-X 1 are reduced with \texttt{OSA v11.0}. We follow the standard steps described in the JEM-X user 
manual\footnote{\url{https://www.isdc.unige.ch/integral/download/osa/doc/11.1/osa_um_jemx/man.html}.}. Spectra are extracted 
in each scw where the source is automatically detected by the software at the image creation and fitting step, and are then computed from 
$\sim$3 to $\sim$34\,keV with 32 logarithmic bins. The individual spectra are then combined with the \texttt{OSA} \textit{spe\_pick} tool. The appropriate ancillary response files (arfs) are produced during the spectral extraction and combined with \textit{spe\_pick}, while the redistribution matrix file (rmf) is rebinned from the instrument characteristic standard rmf with j\_rebin\_rmf. We add 3\,\% systematic error on all spectral channels for each of the stacked spectra, as recommended in the JEM-X user manual, and the spectra are considered between 3 and 25~keV. 

\section{Results}
\label{sec:analysis}

\subsection{Methodology for the spectral analysis}
We use both the \texttt{Interactive Spectral
  Interpretation System} \citep[\texttt{ISIS};][]{Houck2000} and \texttt{XSPEC
  V.12.10} \citep{Arnaud96} packages for the spectral
analysis.

In order to assess the presence, significance, and main properties of the high-energy tail, we first analyze the data with a simple phenomenological model, that is, a thermal Comptonization continuum in combination with a power law extending into the MeV range (Sect.~\ref{sec:phen}). Keeping in mind the suggested origin for the LHS high-energy tail proposed by L11, \citet{Jourdain2012}, and R15, 
we then investigate other possible origins for this component. We use physical models considering the broad-band source spectrum as originating from a single medium composed of a hybrid thermal/nonthermal particle distribution (Sect.~\ref{sec:eqpair} and~\ref{sec:belm}).

\begin{figure*}
    \centering
    \includegraphics[width=0.8\textwidth]{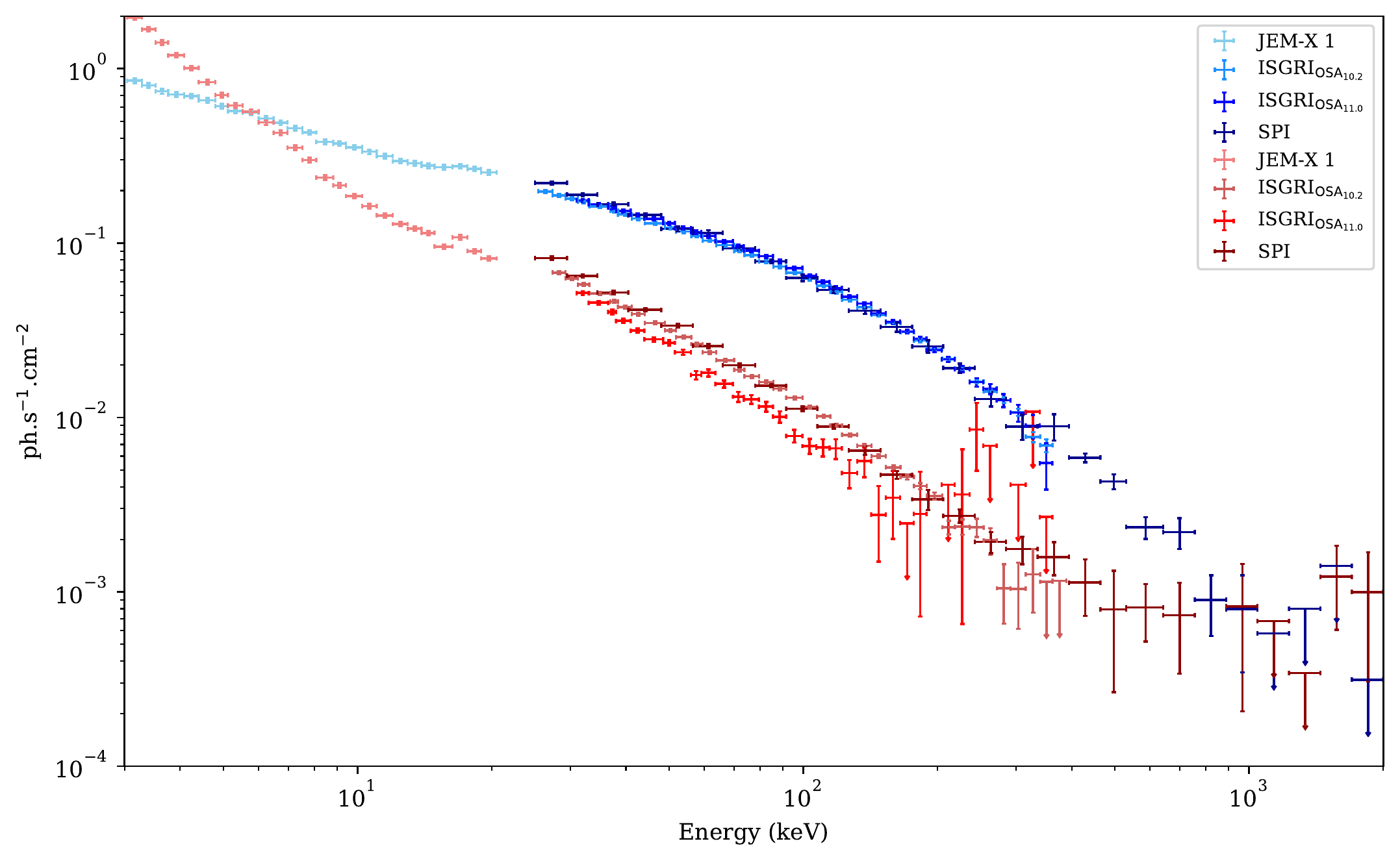}
      \caption{Long-term (2003--2018), broad-band (3 keV to 2 MeV), and state-resolved (red: HSS, blue: LHS) stacked spectra of JEM-X 1 (light colors), ISGRI (intermediate colors), and SPI (darker colors).}
         \label{fig:specres}
\end{figure*}

\subsection{Phenomenological approach: a high-energy tail in both states}
\label{sec:phen}

The LHS and HSS stacked spectra of JEM-X 1, IBIS/ISGRI, and SPI are shown in Fig. \ref{fig:specres}.
Following the approach presented in J14 and R15, we first focus on the 3-400\,keV spectral band. In that energy range, we apply the thermal Comptonization model \texttt{compTT} \citep{Titarchuk1994} being reflected off the disk, which in \texttt{xspec} notation is
\begin{equation*}
    \texttt{const*reflect(compTT)}, 
\end{equation*}
and has been successfully used by J14 and R15. The \texttt{const} component is a multiplicative constant necessitated by the differences in absolute flux calibration between the different instruments used. In the LHS, we freeze the value of the seed photon temperature of the \texttt{comptt} component to $kT_0 = 0.2$\,keV \citep[e.g.,][R15]{Balucinska1991, Wilms2006}. In the HSS, we also need to add the thermal emission from the disk (\texttt{diskbb}) and an iron line at 6.4\,keV (fitted by a gaussian, \texttt{gauss}). Thus, the model is written in \texttt{xspec} as
\begin{equation*}
    \texttt{const*(reflect(compTT)+diskbb+gauss)}. 
\end{equation*}
As the spectral resolution of JEM-X prevents a fine description of the iron line, we fix its energy and its width to 
$E_\mathrm{Fe} = 6.4$ \,keV and $\sigma_\mathrm{Fe} = 0.4$\,keV, respectively. The seed photon temperature of the \
\texttt{comptt} component is fixed to the temperature of the \texttt{diskbb} component. 

\begin{figure}

\includegraphics[width=\columnwidth]{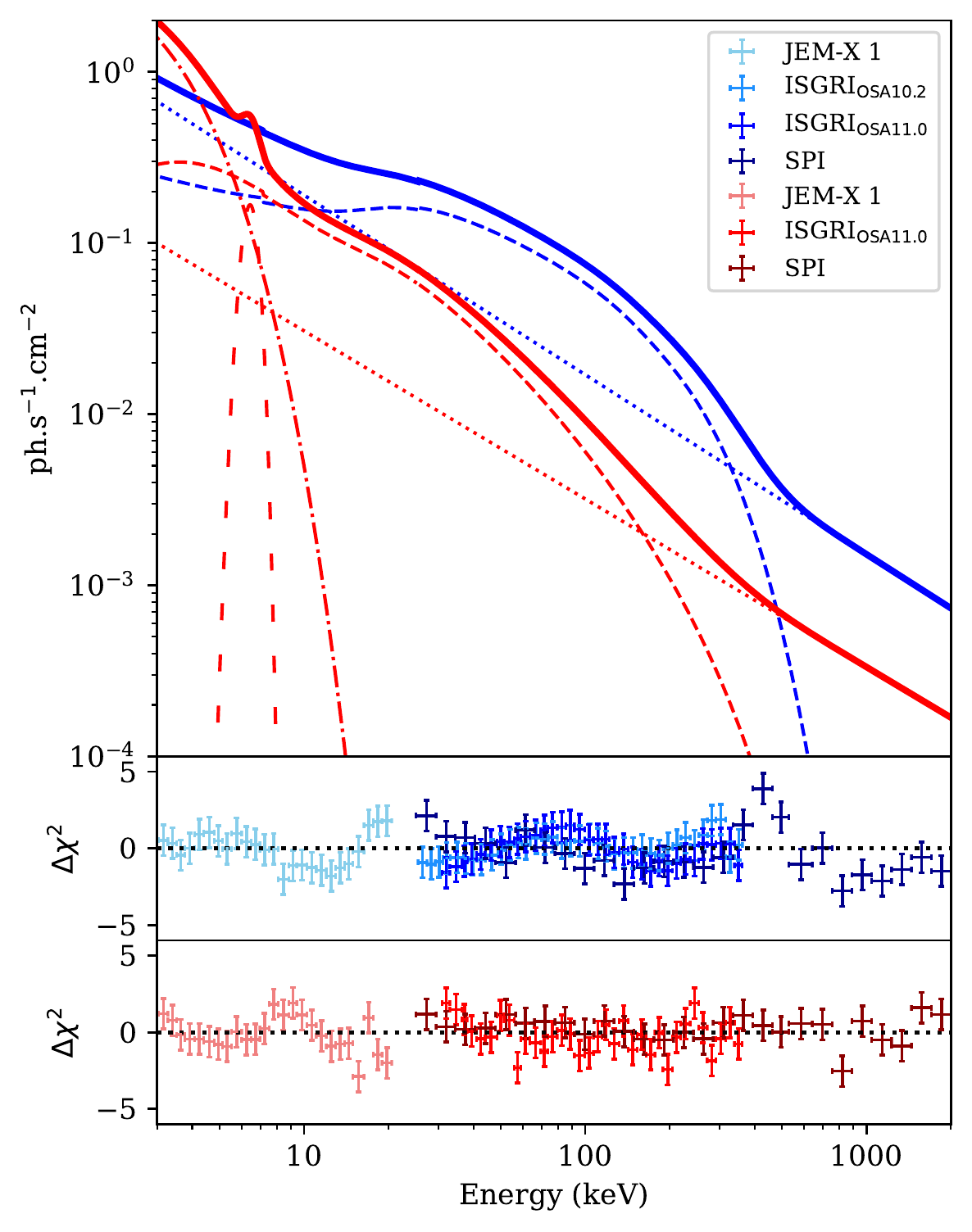}
\caption{Best-fit models obtained from our phenomenological fitting of the 3--2000 keV LHS (blue) and HSS (red) spectra. JEM-X, ISGRI, and SPI data are represented with light, intermediate, and darker colors respectively. The different model components are shown with different line styles: Comptonized continuum  (densely dashed), additional power law (dotted), disk emission (dotted-dashed), and the iron line (loosely dashed). The bottom panels show the residuals between these models and spectra for the source in both states.}
\label{fig:comptt}
\end{figure}

Once an appropriate fit has been obtained with the spectra below 400\,keV, we then add ISGRI and SPI data above 400\,keV and let the parameters vary freely. A clear high-energy excess is visible in both states above the continuum described with the above-mentioned models. We model this additional emission with a power-law component. We assess the significance of the added power-law tail using  an F-test. A low P-value probability of the F-test suggests that the null-hypothesis, that is, the assumption that the thermal  Comptonization continuum describes the broad-band spectrum well, is false and means that the fit requires the additional power-law component to model the high-energy tail. We find an F-test probability of $P = 3.5 \times 10^{-24}$ and $2.6 \times 10^{-4}$ in 
the LHS and HSS, respectively. Table~\ref{table_pars} lists all best-fit parameters in the 3--2000\,keV range and Fig. \ref{fig:comptt} 
shows the best-fit models and their residuals for both states.

\begin{table}
\caption{Model parameters obtained with the phenomenological fits to the LHS and HSS
  3--2000\,keV stacked spectra. Fixed parameters are denoted with an asterisk. }             
\label{table_pars}      
\centering          
\begin{tabular}{l l l}
\hline

Parameters & LHS & HSS \\
\hline
$c_\mathrm{JEM-X}$ & 1$^\ast$ & 1$^\ast$ \\ 
$c_\mathrm{ISGRI,\texttt{OSA10}}$ & $0.85^{+0.04}_{-0.04}$ &  -- \\ 
$c_\mathrm{ISGRI,\texttt{OSA11}}$ & $0.88^{+0.04}_{-0.04}$ & $0.89^{+0.03}_{-0.03}$ \\
$c_\mathrm{SPI}$ & $0.90^{+0.04}_{-0.04}$ & $1.20^{+0.00}_{-0.02}$ \\
$kT_0$ [keV] & 0.2$\ast$ & $0.9^{+0.1}_{-0.1}$ \\
$kT$ [keV]  & $53^{+3}_{-3}$ & $62^{+71}_{-28}$ \\
$\tau$ & $1.47^{+0.09}_{-0.09}$ & $0.29^{+0.02}_{-0.02}$\\
$\Gamma_\mathrm{tail}$ & $2.06^{+0.03}_{-0.03}$& $2.0^{+0.3}_{-0.4}$\\
$\Omega/2\pi$ & $0.8^{+0.2}_{-0.1}$ & $0.5^{+0.2}_{-0.2}$ \\ 
$E_\mathrm{Fe}$ [keV] & -- & 6.4$^\ast$ \\
$\sigma_\mathrm{Fe}$ [keV] & -- & 0.4$^\ast$ \\

F$_{0.4-2\mathrm{\,MeV}}$ & -- & -- \\

[$\times 10^{-9}$\,ergs. cm$^{-2}$. s$^{-1}$] & 3.8 & 0.9 \\

$\chi^2_\mathrm{red}$ & 141.54/117& 103.29/78 \\
\hline

\end{tabular}
\end{table}

\subsubsection{The LHS}
\label{sec:lhs_phen}

In the LHS, the electron temperature, $kT \sim 53$\,keV,  is consistent with the value obtained by J14 and roughly consistent with that of 
R15. The optical depth, $\tau \sim 1.5$,  seems here slightly higher than in the two previous studies. The reflected photon fraction 
of $\Omega/2\pi \sim 0.8$ is consistent with the one found by J14 but remains higher than in R15 (where $\Omega/2\pi = 0.13 \pm 0.02$). 
These small differences are probably due to the intrinsically variable nature of the source, our largely extended 
observations sample and integrated exposure time in this state compared to these previous studies, and the 
extended spectral range we consider here compared to J14 and R15, and are therefore not surprising. As expected from the comparison of our SPI spectrum with that of R15 (Fig.~\ref{fig:spipicsit}), we find a softer photon index ($\Gamma_\mathrm{tail} \sim$ 2.06; Table~\ref{table_pars}) for the additional power-law component than the (very hard) value found by R15 ($\Gamma_\mathrm{tail}^\mathrm{R15}$=$1.4^{+0.2}_{-0.3}$). Our results nevertheless remain rather consistent with those of J14 ($\Gamma_\mathrm{tail}^\mathrm{J14}$=$1.8 \pm 0.2$).

\subsubsection{The hard-soft state}
\label{sec:hss_phen}
Significant residuals arise in the HSS between tens to hundreds of keV for IBIS/ISGRI when extracted with \texttt{OSA v10.2}. We therefore exclude the \texttt{OSA v10.2} HSS data from the fit. The resulting reduction in count statistics has only a minor impact due to the coverage of the same energy range by SPI. 

We are able to integrate a sufficient number of counts in the HSS of Cyg X-1 and can confirm the suggestion previously made by J14
of a possible high-energy tail in the HSS. We find a photon index for the additional power law 
 $\Gamma_\mathrm{tail}\simeq 2.0$ (Table~\ref{table_pars}), which is consistent with the value we find in the LHS. We find an electron temperature of $62^{+71}_{-28}$ , which is not well constrained, and an optical depth of $\tau\simeq 0.29$ consistent with the previous results of J14. We can constrain the reflection component to $\Omega/2\pi \simeq 0.5$ (Table~\ref{table_pars}) . 

\subsection{State-resolved spectral modeling: unmagnetized hybrid plasma}
\label{sec:eqpair}

\subsubsection{Basics of the \eqpair model}
To go beyond the simple phenomenological approach, we model the data with the emission from a single region, where the electron distribution is hybrid. In such a model, the high-energy tail is explained by the emission of a nonthermal population of accelerated particles. These particles cool down and eventually thermalize because of several processes such as Coulomb collisions. The steady-state distribution is then a hybrid distribution constituted by a low-energy, thermal population producing the bulk of the emission below 400 keV and a high-energy, nonthermal population producing the high-energy tail above 400 keV. 

To model such a hybrid plasma, we first use \eqpair \citep{Coppi1999}, which has been successfully used in describing the broad-band spectra of Cyg X-1 in different states \citep{Gierlinski1999, McConnell2002, Wilms2006, Cadolle2006, Fritz2007, Nowak2011, DelSanto2013}. This model takes into account all microphysical interactions and processes between particles and photons (bremsstrahlung, Compton-scattering, pair production, and electron-electron Coulomb collisions) in a homogeneous, isotropic, and ionized medium of total optical depth $\tau_\mathrm{T}$ which is the sum of the proton Thomson optical depth $\tau_\mathrm{p}$ plus the optical depth of electron-positron pairs. Soft photons are injected into the medium as a multicolor disk black-body characterized by the inner disk temperature $\mathrm{k}T_\mathrm{bb}$ and the total power $L_{s}$. Energy is provided to the electron population through two channels: either directly in the thermal pool of electrons with power $L_\mathrm{th}$, or as nonthermal particles with a power-law distribution with index $\Gamma_\mathrm{inj}$ and power $L_\mathrm{nth}$. The total power injected into electrons is $L_\mathrm{h}=L_\mathrm{th}+L_\mathrm{nth}$. In steady state, the source luminosity is equal to the total injected power: $L=L_\mathrm{s}+L_\mathrm{h}$. In the code, the luminosity is described by the compactness parameter:
\begin{equation}
    l = \frac{\sigma_\mathrm{T}}{m_\mathrm{e}c^3} \frac{L}{R}
,\end{equation}
where $\sigma_\mathrm{T}$ is the Thomson cross-section, $m_\mathrm{e}$ the electron mass, and $R$ is the typical source size. Other compactness parameters $l_s$, $l_{h}=l_{th}+l_{nth}$ are defined in a similar manner. The overall spectral shape is strongly dependent on the dominance of high-energy emission, that is, the ratios
$l_\mathrm{h}/l_\mathrm{s}$ and $l_\mathrm{nth}/l_\mathrm{h}$.

\subsubsection{Application to our data}

As in all our different approaches, we freeze the system inclination to 27\degr. In the LHS, we fix the temperature of 
the disk photons to \kTbb\ $=0.2$\,keV and the disk compactness to \ls $=1$ \citep[e.g.,][]{Gierlinski1999, McConnell2002, DelSanto2013}. 
In the HSS, the spectrum is dominated by the emission of the disk photons and therefore, we let the disk temperature, \kTbb, vary freely 
and fix \ls to 10 \citep{Gierlinski1999, McConnell2002, DelSanto2013}. We further assume the Lorentz factor of the nonthermal plasma to be power-law distributed between $\gamma_\mathrm{min} = 1.3$ and $\gamma_\mathrm{max} = 1000$. The radius of the corona is set to $10^{7}$\,cm and we assume solar abundances ANGR \citep{Anders1989} for all metals in the reflection modeling. The ionization parameter is set to zero and we do not inject pairs. As in our previous approach, we also add a Gaussian to the HSS model.

%

\begin{table}
\caption{\eqpair model parameters for our stacked HSS and LHS
  spectra. Fixed parameters are denoted with an asterisk.}
\label{table:spec_par_eqpair}
\centering
\begin{tabular}{lll}
\hline\hline
Parameter & LHS & HSS \\
\hline
$c_\mathrm{JEM-X}$ & 1$^\ast$ & 1$^\ast$ \\ 
$c_\mathrm{ISGRI,\texttt{OSA10}}$ & $0.96^{+0.02}_{-0.02}$ &  -- \\
$c_\mathrm{ISGRI,\texttt{OSA11}}$ & $0.99^{+0.04}_{-0.04}$ & $0.88^{+0.03}_{-0.03}$ \\
$c_\mathrm{SPI}$ & $1.00^{+0.02}_{-0.02}$ & $1.20^{+0.00}_{-0.03}$ \\
\kTbb [eV] & 200$^\ast$ &  $565^{+68}_{-58}$\\
\lhls & $6.34^{+0.30}_{-0.10}$ & $0.42^{+0.05}_{-0.05}$ \\
\ls & 1$^\ast$ & 10$^\ast$ \\
\lnthlh & $0.9^{+0.0}_{-0.2}$ & $>0.8$ \\
\taup & $1.29^{+0.08}_{-0.08}$ & 0.2$^\ast$ \\
\Ginj & $2.9^{+0.2}_{-0.1}$ & $3.1^{+0.1}_{-0.4}$ \\
\refl & $0.25^{+0.01}_{-0.03}$ & $1.0^{+0.6}_{-0.2}$ \\
$E_\mathrm{Fe}$ [keV] & -- & 6.4$^\ast$ \\
$\sigma_\mathrm{Fe}$ [keV] & -- & 0.4$^\ast$ \\

$\chi^2/ \mathrm{dof}$ & $134.68/117$ & $101.81/79$ \\
\hline
\end{tabular}
\end{table}

We show the best-fit model in Fig. \ref{fig:eqair_belm} and list all parameters in Table~\ref{table:spec_par_eqpair}. Our parameters are consistent overall  with the previous studies indicated above.  

The LHS is clearly dominated by Comptonization at an optical depth of $\sim 1.3$ and with a high fraction of accelerated nonthermal particles with $l_\mathrm{nth}/l_\mathrm{h} \sim 0.9$. This results in a spectrum that is dominated by the hard X-rays, 
that is, $l_\mathrm{h}/l_\mathrm{s}\sim 6.3$. The accelerated electron distribution is a power law with an injection electron index 
$\Gamma_\mathrm{inj}\simeq 2.9$ (Table~\ref{table:spec_par_eqpair}).

Our HSS spectra show some emission up to 600\,keV, i.e., to slightly higher energies than in most previous studies. In this state, our fit becomes unstable when letting $\tau_\mathrm{p}$ vary freely. We therefore freeze this parameter to $\tau_\mathrm{p}$= 0.2 in line with the values previously obtained with \eqpair \citep{Gierlinski1999, Frontera2001, McConnell2002, Sabatini2013, DelSanto2013} and in coherence with our phenomenological approach (Sec.~\ref{sec:hss_phen}; Table~\ref{table_pars}). We note that we need an 
 injection electron index of $\Gamma_\mathrm{inj}\sim$ 3.0 consistent with the value we find in the 
 LHS (Table~\ref{table:spec_par_eqpair}). The spectrum is dominated by the  disk thermal emission as indicated by a 
 \lhls$\sim$0.4 fraction weaker and a photon disk temperature of \kTbb$\sim$0.6\,keV, which is higher than in the LHS and is consistent with the results of \citet{Gierlinski1999}, \citet{Frontera2001}, \citet{McConnell2002}, \citet{Sabatini2013}, \citet{DelSanto2013}. The fraction of particles emitted by nonthermal processes \lnthlh is consistent with a purely nonthermal distribution ($> 77$\,\%). Finally, the reflected photon fraction is consistent with our previous approach, although it is less constrained.

\subsection{State-resolved spectral modeling: magnetized hybrid plasma}
\label{sec:belm}

\subsubsection{Basics of the \belm model}
In the previous section, we model the emission from an unmagnetized medium with the \eqpair model. It is nevertheless believed that the disk and the jet are significantly magnetized \citep[e.g.,][]{Blandford1977, Ferreira2006}. Moreover, nonthermal particles are often assumed to be accelerated by magnetic reconnection, which also requires strong magnetic fields \citep[e.g.,][]{Poutanen2009}. 

In this section we therefore extend the previous modeling by additionally considering magnetic fields as a natural explanation for the accelerated, nonthermal particle distribution that we used for \eqpair.  To do so, we use the \belm code \citep{Belmont2008}, which uses the same parameters as \eqpair, but additionally introduces the magnetic field intensity through the magnetic compactness,
\begin{equation}
\centering
    l_\mathrm{B} = \frac{\sigma_\mathrm{T}}{m_\mathrm{e}c^2}R\frac{B^2}{8\pi}.
    \label{eq:lb}
\end{equation}
The inclusion of the magnetic field yields several effects over the modeling presented so far. Firstly, additional synchrotron radiation is produced, which can then be Comptonized (synchrotron self-Compton). This directly impacts the spectral shape by adding a new broad-band component and indirectly by increasing the particle cooling rate. Secondly, low-energy particles can synchrotron-absorb high-energy photons, which heats these particles, resulting in efficient thermalization. 

To moderate the model degeneracy, we focus on a purely nonthermal model \citep[\lth $=0$, see also][]{DelSanto2013}. Thus, the total radiation compactness $l = \ls + \lth +\lnth$ becomes $l = \ls + \lnth$, and $l_\mathrm{h} =\lnth$. 

The table we use is computed with five model parameters: \lnth, \lBnth, \taup, \Ginj and \kTin. As shown in \cite{Malzac2009}, the shape of the spectrum depends on ratios such as $\lBnth$ or $\lnth/l$ and not on the value of the total compactness $l$. Therefore, its value is fixed to $l = 10$ in the table which represents the typical value of $0.01 L_\mathrm{Edd}$ for both states, while $\ls$ is constrained by enforcing $\ls + \lnth =10$. The exact magnetic field compactness depends on the normalization $N$ of the model,
\begin{equation}
    l_{B,\mathrm{obs}} = \frac{l_B}{l_\mathrm{nth}}\frac{l_\mathrm{nth}}{l} l_\mathrm{obs} \quad \text{with} \quad  l_\mathrm{obs} = \frac{4 \pi D^2 \sigma_T}{R c} N l.
    \label{eq:lbobs}
\end{equation}
The magnetic field intensity can then be estimated assuming the distance to the source, $D = 1.86$\,kpc, and the size of the corona, $R = 20\,R_\mathrm{g}$. As previously, we fix the value of the injected seed photons to $kT_\mathrm{in}  = 0.2$\,keV in the LHS and $\tau_\mathrm{p} =0.2$ in the HSS. An iron line fixed at 6.4\,keV is also added in the HSS. Here, we describe reflection using the convolution code \texttt{reflect}. In \eqpair, the reflection model \texttt{ireflect} is directly implemented as part of the model, with the additional capability to model an ionized medium. In our case, we choose a neutral medium in order to guarantee comparability between \belm and \eqpair.

\subsubsection{Application to our data}

\begin{figure*}[h]
\centering
\includegraphics[width=\textwidth]{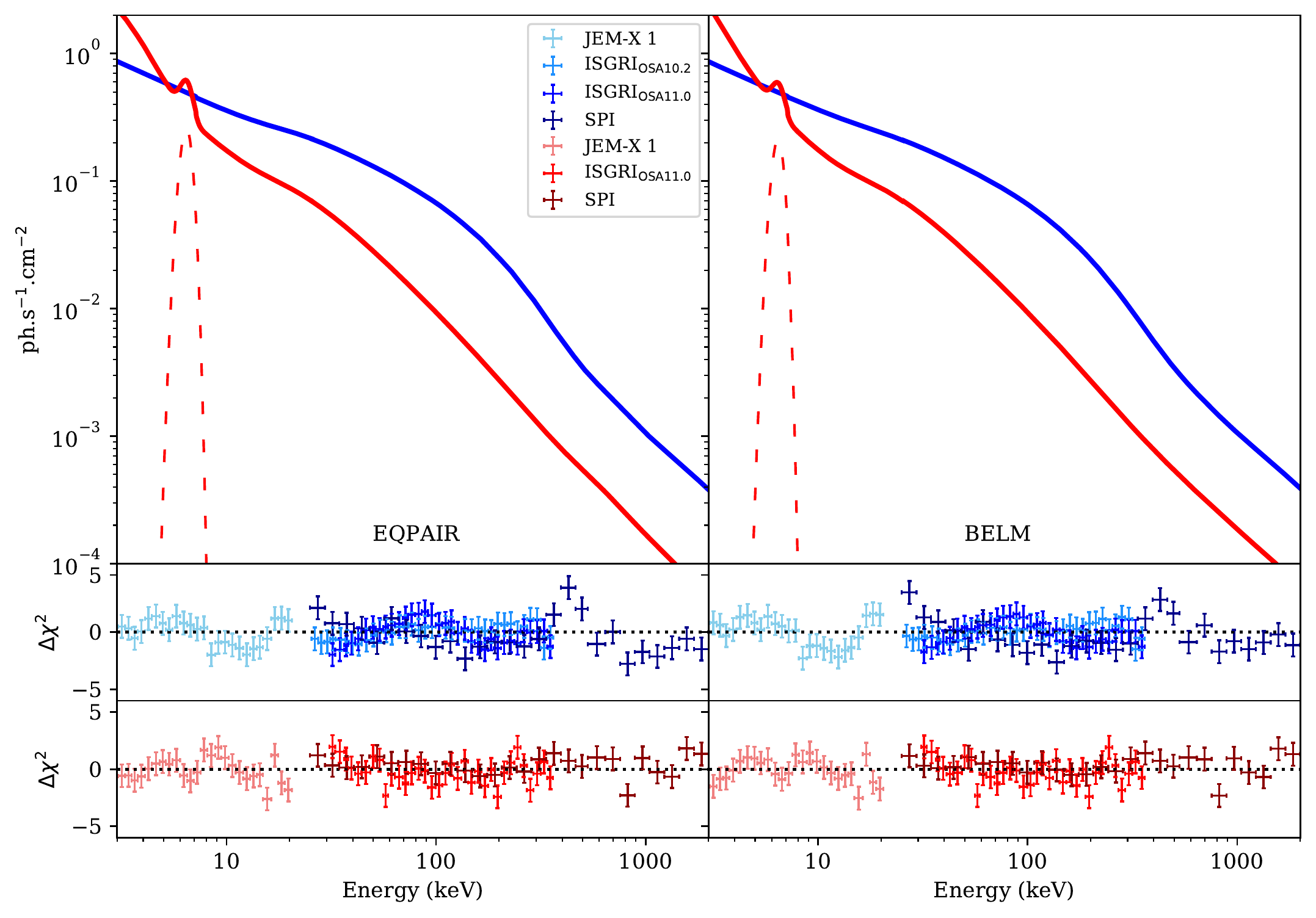}

\caption{Best models obtained from the fits to spectra for Cyg X-1  with the unmagnetized (left) and magnetized (right) hybrid corona models. Plain lines correspond to the full models whereas the loosely dashed line shows the iron line component. JEM-X, ISGRI, and SPI data are represented in light, intermediate, and darker colors, respectively.}
\label{fig:eqair_belm}
\end{figure*}

Table~\ref{table:spec_par_belm} shows the best-fitting parameters for our modelings with \belm. The best-fit models 
and the residuals between those and the source spectra are shown in Fig.\ref{fig:eqair_belm}.

The LHS is only compatible with weak values of the magnetic compactness parameter (\lBnth $<0.04$). This results in a coronal 
magnetic field  $B < 4.83 \times 10^{5} (20 R_\mathrm{g}/R)$\,G  compatible with the results obtained by 
\citet{Malzac2009}, \citet{DelSanto2013}, and \cite{Poutanen2009}. Larger fields produce large amounts of soft synchrotron photons that 
Compton cool the thermal electron population. The parameters obtained in this state (Table~\ref{table:spec_par_belm}) 
are consistent with the parameters found with the unmagnetized model \eqpair (Table~\ref{table:spec_par_eqpair}), except for the optical 
depth for which we observe a larger value with \belm. The difference in optical depths may arise from a different treatment of the
 coronal geometry: \belm uses an escape probability corresponding to a pure spherical geometry, while \eqpair uses a combination 
 of factors related to spherical and slab geometries. 

In the HSS, we also find parameters consistent with the unmagnetized model. The parameters are compatible with the previous results of \citet{Malzac2009}, \citet{Poutanen2009}, \citet{DelSanto2013}. We note that $l_\mathrm{B}$ and \ls both contribute to the corona cooling. \ls is by definition the total amount of cold photons entering in the corona, and thus, the fraction of disk photons that participate in the cooling is 100\,\%. We also test the possibility that a fraction of the disk photons will not interact with the corona by adding a disk black body (\texttt{diskbb}) in the model and investigate whether it has an effect on the value of \lBnth and/or \lnth. The seed photon temperature of \texttt{diskbb} is fixed to the temperature $kT_\mathrm{in}$ in the \belm model. We find that the uncertainties were slightly different: \lBnth$= 1.6 \pm 0.4$ and $l_\mathrm{nth} = 2.2^{+0.9}_{-0.2}$ with the addition of \texttt{diskbb} compared to \lBnth $= 1.6^{+0.4}_{-0.7}$ and $l_\mathrm{nth} = 2.2^{+0.7}_{-0.2}$ without (Table \ref{table:spec_par_belm}). This latest value on \lBnth leads to a magnetic field of $B = 1.9^{+1.6}_{-1.3} \times 10^{6} (20 R_\mathrm{g}/R$)\,G. 

The shape of the Comptonization spectrum depends on the heating/cooling ratio; a smaller \ls and a larger $l_\mathrm{B}$ can lead to the same Comptonization spectrum. However, contrary to a disk, the thin synchrotron spectrum extends above 1\,keV; in the LHS, this component can be hidden by the Comptonized component, but in the HSS, if there was a synchrotron emission comparable to the disk emission, we should see it in the spectrum, which is not the case and this is confirmed by the fits with an additional disk black-body component.
We discuss these results and their implications in the following section.

%
%
%

\begin{table}[h]
\caption{\belm model parameters for our stacked LHS and HSS  spectra. Fixed parameters are denoted with an asterisk.}

\label{table:spec_par_belm}
\centering

\begin{tabular}{llll}
\hline\hline
Parameter & LHS & HSS \\
\hline
$c_\mathrm{JEM-X}$ & 1$^\ast$ &  1$^\ast$\\
$c_\mathrm{ISGRI,\texttt{OSA10}}$ & $0.96^{+0.03}_{-0.03}$ &  --\\
$c_\mathrm{ISGRI,\texttt{OSA11}}$ & $0.99^{+0.03}_{-0.03}$ & $0.88^{+0.03}_{-0.03}$ \\
$c_\mathrm{SPI}$ & $1.02^{+0.03}_{-0.03}$ & $1.20^{+0.00}_{-0.01}$ \\

\kTin\,[eV] & 200$^\ast$ & $483^{+67}_{-26}$ \\
\lnth & $9.0^{+0.6}_{-0.1}$ & $2.2^{+0.7}_{-0.2}$ \\
\lBnth & $<0.04$ & $1.6^{+0.4}_{-0.7}$ \\
\taup & $2.08_{-0.09}^{+0.08}$ & 0.2$^\ast$ \\
\Ginj & $3.0^{+0.2}_{-0.1}$ &  $2.4^{+0.3}_{-0.4}$ \\
\refl & $0.16^{+0.03}_{-0.03}$ &  $1.1^{+0.4}_{-0.2}$ \\
$\chi^2/ \mathrm{dof}$ & $141.61/117$ & $ 97.73/79$ \\
\hline
\end{tabular}
\end{table}

\section{Discussion}
\label{sec:discussion}

\subsection{Summary of the main results}
    
The extended \integral data set considered here allow us to confirm the detection of a high-energy tail in the LHS \citep[e.g.,][L11, J14, R15]{Cadolle2006}. We also report the detection of a high-energy tail in the HSS with \integral, extending to at least 600\,keV as seen with SPI (Fig.~\ref{fig:specres}). 

Our empirical model (Sect.~\ref{sec:phen}) consists of purely thermal Comptonization (\texttt{compTT}) with reflection off neutral gas and an independent power-law component to model any high-energy tail emission. We bring solid constraints for the high-energy tails in both states: with $\Gamma^\mathrm{LHS}\simeq 2.03$ and $\Gamma^\mathrm{HSS} \simeq 2.0$ (Table~\ref{table_pars}) they, in particular, appear to have  
similar slopes. Their 200--1000 keV fluxes are $3.5 \times 10^{-9}$ and $7.1 \times 10^{-10}$\,ergs/cm$^2$/s which are somewhat different.

In an attempt to gain a more physical description of the data, we tested hybrid Comptonization models for an unmagnetized (\eqpair) and magnetized (\belm) corona. Our results are consistent with those of previous studies \citep{Gierlinski1999, McConnell2002, DelSanto2013}. The fits with  
 \eqpair lead to a higher value of \lhls in the LHS than in the HSS, which indicates that the source emission is dominated by Comptonization in the LHS. Both states are well modeled and we obtain electron injection indices of $\Gamma^\mathrm{LHS}_\mathrm{inj}\simeq 2.9$ and $\Gamma^\mathrm{HSS}_\mathrm{inj}\simeq 3.1$. These results are globally consistent with the parameters we obtain with \belm. Moreover, \belm provides us constraints on 
 the magnetic field. We found an upper limit $B < 4.83 \times 10^{5} (20 R_\mathrm{g}/R)$\,Gauss in the LHS, and we are able to constrain $B = 1.9^{+1.6}_{-1.3} \times 10^{6} (20 R_\mathrm{g}/R)$\,Gauss in the HSS. Those values are consistent with upper limits found by \cite{Malzac2009}, \cite{Poutanen2009}, \cite{DelSanto2013}.

\subsection{Interpretation: origin of the high-energy emission}

The detection of a high-energy tail in both states leads us to consider two different scenarios:

\begin{itemize}
\item [1.] The high-energy tails in the LHS and HSS originate from the same medium.
\item [2.] The high-energy tails in the LHS and HSS are the signatures of different physical mechanisms, and thus have distinct origins, 
despite having photon indices of  similar values.
\end{itemize}

\subsubsection{Same origin and spectral interpretation}

Our results with the semi-phenomenological approach are consistent with what has been previously reported for Cyg X-1: 
a significantly higher optical depth in the LHS compared to the HSS, which could suggest that the corona shrinks as the system transits from the LHS to HSS \citep[e.g.,][]{Zdziarski1998,Wilms2006,DelSanto2013}, and an electron temperature around 50\,keV in the LHS. However, this approach does  not allow us to treat the broadband spectra in a self-consistent way, and prevents us from verifying whether or not a single medium (a corona) can be responsible for the 100--1000\,keV emission.

With the physical, broad-band hybrid spectral models \eqpair and \belm, in contrast, we can study the evolutionary behavior of a corona as a single self-consistent medium. The first conclusion we can draw is that our spectra are well modelled with both the unmagnetized and magnetized models \eqpair and \belm. In these frameworks, the high-energy emission is produced by a nonthermal electron population in either a nonmagnetized or a magnetized corona in both states.

As \belm furnishes precious constraints on the magnetic field in both states, we can, moreover, draw another conclusion which appears intriguing. The ratio between the equipartition magnetic field compactness $l_\mathrm{B_R}$ and the magnetic field compactness $l_\mathrm{B}$ is given by \cite{Lightman1987}:

\begin{equation}
\frac{l_\mathrm{B}}{l_\mathrm{B_R}} = \frac{l_\mathrm{B}}{l_\mathrm{nth}} \frac{l_\mathrm{nth}}{l} \frac{4\pi/3}{1+\tau_\mathrm{T}/3}
.\end{equation}
We find $\frac{l_\mathrm{B}}{l_\mathrm{B_R}} < 0.08$ and $1.4^{+0.7}_{-0.8}$ for the LHS and HSS, respectively. In the LHS, this low value argues against magnetic reconnection powering the corona, which is consistent with the conclusions of \citet{Malzac2009}, \citet{Poutanen2009}, and \citet{DelSanto2013}. If electron acceleration is not possible via magnetic reconnection, it thus implies that the high-energy emission must be produced in a second, independent zone, which is not the corona, and could for example be the jets. 

In the framework of \belm as a one-zone model, the MeV tail is mainly due to Comptonization and would be inconsistent with the observed high degree of polarization claimed for the LHS so far \cite[L11, ][R15]{Jourdain2012}. In the HSS, the magnetic field is large enough to be able to produce magnetic reconnection, and this thus could be the mechanism responsible for the electron acceleration in the corona \citep{Malzac2009, Poutanen2009, DelSanto2013}.
To summarize, an unmagnetized corona could be at the origin of the high-energy emission in both states and is thus a 
plausible scenario for both, while a magnetized corona seems to be working only in the HSS.

\subsubsection{Different origin and comparison with previous results}

As indicated, the tails in both long-lasting spectral states do not necessarily have the same origin. Evolved radio jets \citep[e.g.,][and references therein]{Rushton2012}, the high polarization fraction of the high-energy tail above $400$\,keV \citep[L11, ][R15]{Jourdain2012}, and broadband high-energy spectral features associated with the jets of  \cyg (L11, \citealt{Jourdain2012}, \citealt{Zdziarski2012}, R15, \citealt{Walter2017}) seem to point towards a synchrotron jet emission in the LHS. Several studies investigate this possibility. \cite{Zdziarski2012} apply their jet model and find that such emission requires an electron acceleration at a rather hard power-law index (1.3--1.6), which is difficult to reconcile with standard acceleration mechanisms. Similar conclusions were found in an another source observed with \integral, namely GRS 1716--246, for which \cite{Bassi2020} used a jet internal shock emission model \citep{Malzac2013} in order to model the high-energy emission of this source. However, another study on Cyg X-1 shows that the emission during the LHS can be explained by a self-consistent jet model \citep{Kantzas2020}. In their model, a broadband jet-synchrotron component is the origin of both the radio continuum and the MeV tail in the LHS. A photon index of $\Gamma \sim 2$ in the MeV range would be compatible with their results. The jet origin is further supported by the detection between 60\,MeV and 20\,GeV by \textit{Fermi}/LAT \citep{Bodaghee2013,Malyshev2013,Zanin2016,Zdziarski2017}, which may also hint towards a hadronic contribution to the jet-accelerated plasma \citep{Kantzas2020}.


A scenario that lies beyond the scope of our paper, and is hard to constrain and fully discuss with our data, is the putative
presence of a jet in both states. This possibility arises from recent claims of possible detected jets in the HSS. In particular, shortly  after the LHS-to-HSS state transition around MJS~55400, \citet{Rushton2012} detect a weak and compact jet on VLBI scales. It is unclear whether this feature has remained undetected in the long-term stable HSS before, or whether this is a remnant feature of the state transition that happened close-by in time. A recent study claims the detection of radio jets in the HSS of Cyg X-1 \citep{Zdziarski2020}. The authors observe a clear correlation between 15\,GHz radio flux and the simultaneous 15--50\,keV flux not only in the LHS but also, albeit with a different slope, in the HSS. However, as opposed to the LHS, the soft X-rays at a few keV are reported to be time-lagged by about 100\,days in the HSS, suggesting advection from the donor through the accretion disk. The absence of a time-lag between the radio and hard (15--50\,keV) X-ray emission in both states might, here, imply a jet origin for the latter hard X-ray radiation. However, we note that the significantly lower degree of polarization below $\sim 400$\,keV \citep{Jourdain2012} and both our spectral modeling and that of \cite{Kantzas2020} suggest that the persistent jet may only contribute above hundreds of keV during the LHS. 

No radio emission has been detected in any other microquasar in the HSS down to a very low flux level \citep[e.g.,][]{Fender99, Remillard2006, Belloni2010}. Still, a recent study suggests that jets could survive in the HSS with heavily suppressed radio emission that may or may not be detectable depending on the observing technique and intrinsic radio-flux level. These so-called ``dark jets'' are kinematically dominated and radiatively inefficient \citep{Drappeau2017}. In the context of the internal shock model of \citet{Drappeau2017}, the radio flux is quenched, but a kinematic jet driven by the disk rather than the corona still subsists. However, this study is based on a broadband modeling of GX~339$-$4 and H~1743$-$322, both of which are microquasars with low-mass companion stars and are purely Roche lobe accreting. It is unclear whether or not the results from \cite{Drappeau2017} can be applied to \cyg, a persistently wind-accreting system with a high-mass companion. 

  
\section{Conclusion}
\label{sec:conclusion}    
In this paper, we advance the long-lasting investigation of the nature of the hard MeV tail of \cyg for both major spectral states. We accumulate data from 15\,years of \integral (JEM-X, IBIS and SPI) observations of the source and reach unprecedented count statistics in the 3-2000\,keV range, not only  in the LHS, but also in the HSS, which allows us to firmly confirm the suspected presence of a high-energy component ---extending to at least 600 keV-- in the spectra of the source. 

We follow two approaches with the modeling of these data.
In the first, empirical model, we fit a thermal Comptonization spectrum independent of a simple power law, describing the MeV tail between hundreds of keV and $\sim$1\,MeV. The slopes of the MeV tails appear compatible between the LHS and the HSS. The origin of this high-energy emission can be tackled with single physical spectral models where the data are self-consistently considered. We therefore used the hybrid thermal/nonthermal Comptonization codes \eqpair and \belm, accounting for an unmagnetized and magnetized single-zone corona, respectively. We reasonably successfully reproduce the spectral changes across the states of Cyg X-1 overall. Within this single-zone approach, an unmagnetized corona is able to account for the high-energy emission in both states. Considering the magnetized coronal model, in the LHS, the magnetic field is too weak to be able to accelerate electrons via magnetic reconnection. In the HSS, this mechanism could be responsible for the nonthermal emission we observe \citep{Malzac2009, Poutanen2009, DelSanto2013}. 

We then also discuss our results in the context of physical broadband multi-zone models from the literature and their ability to 
reproduce the radio-to-gamma-ray emission of \cyg. In the LHS, a persistent radio jet can explain both the radio emission and the MeV tail via a direct synchrotron component, while the X-rays at intermediate energies are well described with hybrid Comptonization. When switching to the HSS, we suggest that the persistent jet may transition to a quenched or failed jet, leaving behind a hybrid corona that could either be magnetized or not, and which is responsible of the nonthermal emission. 


Our study does not provide conclusive evidence in favor of any particular one of the scenarios proposed to explain the origin of the high-energy tail in the HSS. A polarimetric analysis of the MeV tails for both states is in preparation, extending the work previously presented in R15. Polarimetry will be key to breaking degeneracies arising between the different physical models, but is currently still affected by large uncertainties. Continued observations and further studies of Cyg X-1, especially in the HSS at energies beyond 1\,MeV with \integral, will be crucial to increasing the count statistics and thus to better constraining the slope of the MeV tail. Complementary studies of other sources would facilitate our understanding of universal properties and differences in such high-energy tails beyond \cyg.

\begin{acknowledgements}
We thank D.~Kantzas, M.~Lucchini, A.~Chhotray, and S.~Markoff for
extensive discussions on the origin of the high-energy tail emission based on physical jet models developed at the Anton Pannekoek Institute for Astronomy, Univ. of Amsterdam.  FC, JR, PL acknowledge partial funding from the French Space Agency (CNES), and the French programme national des hautes \'energies (PNHE). FC and VG acknowledge support from the ESTEC Faculty Visiting Scientist Programme. TS is supported by the German Research Society (DFG-Forschungsstipendium SI 2502/1-1). VG is supported through the Margarete von Wrangell fellowship by the ESF and the Ministry of Science, Research and the Arts Baden-W\"urttemberg. This work was partly funded by Deutsches  Zentrum f\"ur Luft- und Raumfahrt grant 50\,OR\,1411 and 50\,OR\,1909. We made use of ISIS functions provided by ECAP/Remeis Observatory and MIT (\url{http://www.sternwarte.uni-erlangen.de/isis/}). We thank J.E.~Davis for the development of the \texttt{slxfig} module that has been used to prepare the figures in this work. This work is based on observations with \textsl{INTEGRAL}, an ESA project with instruments and science data centre funded by ESA member states (especially the PI countries: Denmark, France, Germany, Italy, Switzerland, Spain), and with the participation of the Russian Federation and the USA. This research has made use of NASA's Astrophysics Data System Bibliographic Services (ADS).
\end{acknowledgements}


\bibliographystyle{aa}
\bibliography{aa_abbrv,mnemonic,bib_CygX-1}

\begin{thebibliography}{66}
\expandafter\ifx\csname natexlab\endcsname\relax\def\natexlab#1{#1}\fi

\bibitem[{{Akaike}(1974)}]{Akaike1974}
{Akaike}, H. 1974, IEEE Transactions on Automatic Control, 19, 716

\bibitem[{{Anders} \& {Tobin}(1989)}]{Anders1989}
{Anders}, A.~K. \& {Tobin}, R.~C. 1989, Journal of Applied Physics, 66, 2794

\bibitem[{{Arnaud}(1996)}]{Arnaud96}
{Arnaud}, K.~A. 1996, in Astronomical Society of the Pacific Conference Series,
  Vol. 101, Astronomical Data Analysis Software and Systems V, ed. G.~H.
  {Jacoby} \& J.~{Barnes}, 17

\bibitem[{{Balucinska} \& {Hasinger}(1991)}]{Balucinska1991}
{Balucinska}, M. \& {Hasinger}, G. 1991, \aap, 241, 439

\bibitem[{{Bassi} {et~al.}(2020){Bassi}, {Malzac}, {Del Santo}, {Jourdain},
  {Roques}, {D'A{\`\i}}, {Miller-Jones}, {Belmont}, {Motta}, {Segreto},
  {Testa}, \& {Casella}}]{Bassi2020}
{Bassi}, T., {Malzac}, J., {Del Santo}, M., {et~al.} 2020, \mnras, 494, 571

\bibitem[{{Belloni}(2010)}]{Belloni2010}
{Belloni}, T.~M. 2010, {States and Transitions in Black Hole Binaries}, ed.
  T.~{Belloni}, Vol. 794, 53

\bibitem[{{Belmont} {et~al.}(2008){Belmont}, {Malzac}, \&
  {Marcowith}}]{Belmont2008}
{Belmont}, R., {Malzac}, J., \& {Marcowith}, A. 2008, \aap, 491, 617

\bibitem[{{Blandford} \& {Znajek}(1977)}]{Blandford1977}
{Blandford}, R.~D. \& {Znajek}, R.~L. 1977, \mnras, 179, 433

\bibitem[{{B{\"o}ck} {et~al.}(2011){B{\"o}ck}, {Grinberg}, {Pottschmidt},
  {Hanke}, {Nowak}, {Markoff}, {Uttley}, {Rodriguez}, {Pooley}, {Suchy},
  {Rothschild}, \& {Wilms}}]{Boeck2011}
{B{\"o}ck}, M., {Grinberg}, V., {Pottschmidt}, K., {et~al.} 2011, \aap, 533, A8

\bibitem[{{Bodaghee} {et~al.}(2013){Bodaghee}, {Tomsick}, {Pottschmidt},
  {Rodriguez}, {Wilms}, \& {Pooley}}]{Bodaghee2013}
{Bodaghee}, A., {Tomsick}, J.~A., {Pottschmidt}, K., {et~al.} 2013, \apj, 775,
  98

\bibitem[{{Bouchet} {et~al.}(2003){Bouchet}, {Jourdain}, {Roques}, {Mandrou},
  {von Ballmoos}, {Boggs}, {Caraveo}, {Cass{\'e}}, {Cordier}, {Diehl},
  {Durouchoux}, {von Kienlin}, {Knodlseder}, {Jean}, {Leleux}, {Lichti},
  {Matteson}, {Sanchez}, {Schanne}, {Schoenfelder}, {Skinner}, {Strong},
  {Teegarden}, {Vedrenne}, \& {Wunderer}}]{Bouchet2003}
{Bouchet}, L., {Jourdain}, E., {Roques}, J.~P., {et~al.} 2003, \aap, 411, L377

\bibitem[{{Bouchet} {et~al.}(2011){Bouchet}, {Strong}, {Porter}, {Moskalenko},
  {Jourdain}, \& {Roques}}]{Bouchet2011_diffuseCR}
{Bouchet}, L., {Strong}, A.~W., {Porter}, T.~A., {et~al.} 2011, \apj, 739, 29

\bibitem[{{Cadolle Bel} {et~al.}(2006){Cadolle Bel}, {Sizun}, {Goldwurm},
  {Rodriguez}, {Laurent}, {Zdziarski}, {Foschini}, {Goldoni}, {Gouiff{\`e}s},
  {Malzac}, {Jourdain}, \& {Roques}}]{Cadolle2006}
{Cadolle Bel}, M., {Sizun}, P., {Goldwurm}, A., {et~al.} 2006, \aap, 446, 591

\bibitem[{{Cangemi} {et~al.}(2021){Cangemi}, {Rodriguez}, {Grinberg},
  {Belmont}, {Laurent}, \& {Wilms}}]{Cangemi2021}
{Cangemi}, F., {Rodriguez}, J., {Grinberg}, V., {et~al.} 2021, \aap, 645, A60

\bibitem[{{Cavanaugh} {et~al.}(1997){Cavanaugh}, {Lee}, {Anthony}, \&
  {Mohan}}]{Cavanaugh1997}
{Cavanaugh}, D., {Lee}, S.~A., {Anthony}, L., \& {Mohan}, V. 1997, in APS
  Meeting Abstracts, APS Meeting Abstracts, D.01

\bibitem[{{Chauvin} {et~al.}(2018){Chauvin}, {Flor{\'e}n}, {Friis}, {Jackson},
  {Kamae}, {Kataoka}, {Kawano}, {Kiss}, {Mikhalev}, {Mizuno}, {Ohashi},
  {Stana}, {Tajima}, {Takahashi}, {Uchida}, \& {Pearce}}]{Chauvin2018}
{Chauvin}, M., {Flor{\'e}n}, H.-G., {Friis}, M., {et~al.} 2018, Nature
  Astronomy, 2, 652

\bibitem[{{Coppi}(1999)}]{Coppi1999}
{Coppi}, P.~S. 1999, in Astronomical Society of the Pacific Conference Series,
  Vol. 161, High Energy Processes in Accreting Black Holes, ed. J.~{Poutanen}
  \& R.~{Svensson}, 375

\bibitem[{{Del Santo} {et~al.}(2013){Del Santo}, {Malzac}, {Belmont},
  {Bouchet}, \& {De Cesare}}]{DelSanto2013}
{Del Santo}, M., {Malzac}, J., {Belmont}, R., {Bouchet}, L., \& {De Cesare}, G.
  2013, \mnras, 430, 209

\bibitem[{{Diehl} {et~al.}(2018){Diehl}, {Siegert}, {Greiner}, {Krause},
  {Kretschmer}, {Lang}, {Pleintinger}, {Strong}, {Weinberger}, \&
  {Zhang}}]{Diehl2018_BGRDB}
{Diehl}, R., {Siegert}, T., {Greiner}, J., {et~al.} 2018, \aap, 611, A12

\bibitem[{{Drappeau} {et~al.}(2017){Drappeau}, {Malzac}, {Coriat}, {Rodriguez},
  {Belloni}, {Belmont}, {Clavel}, {Chakravorty}, {Corbel}, {Ferreira},
  {Gandhi}, {Henri}, \& {Petrucci}}]{Drappeau2017}
{Drappeau}, S., {Malzac}, J., {Coriat}, M., {et~al.} 2017, \mnras, 466, 4272

\bibitem[{{Fender} {et~al.}(1999){Fender}, {Corbel}, {Tzioumis}, {McIntyre},
  {Campbell-Wilson}, {Nowak}, {Sood}, {Hunstead}, {Harmon}, {Durouchoux}, \&
  {Heindl}}]{Fender99}
{Fender}, R., {Corbel}, S., {Tzioumis}, T., {et~al.} 1999, \apjl, 519, L165

\bibitem[{{Fender}(2001)}]{Fender01}
{Fender}, R.~P. 2001, \mnras, 322, 31

\bibitem[{{Ferreira} {et~al.}(2006){Ferreira}, {Petrucci}, {Henri},
  {Saug{\'e}}, \& {Pelletier}}]{Ferreira2006}
{Ferreira}, J., {Petrucci}, P.~O., {Henri}, G., {Saug{\'e}}, L., \&
  {Pelletier}, G. 2006, \aap, 447, 813

\bibitem[{{Fritz} {et~al.}(2007){Fritz}, {Wilms}, {Pottschmidt}, {Nowak},
  {Kendziorra}, {Kirsch}, {Kreykenbohm}, \& {Santangelo}}]{Fritz2007}
{Fritz}, S., {Wilms}, J., {Pottschmidt}, K., {et~al.} 2007, in ESA Special
  Publication, Vol. 622, The Obscured Universe. Proceedings of the VI INTEGRAL
  Workshop, 341

\bibitem[{{Frontera} {et~al.}(2001){Frontera}, {Palazzi}, {Zdziarski},
  {Haardt}, {Perola}, {Chiappetti}, {Cusumano}, {Dal Fiume}, {Del Sordo},
  {Orland ini}, {Parmar}, {Piro}, {Santangelo}, {Segreto}, {Treves}, \&
  {Trifoglio}}]{Frontera2001}
{Frontera}, F., {Palazzi}, E., {Zdziarski}, A.~A., {et~al.} 2001, \apj, 546,
  1027

\bibitem[{{Fuchs} {et~al.}(2003){Fuchs}, {Rodriguez}, {Mirabel}, {Chaty},
  {Rib{\'o}}, {Dhawan}, {Goldoni}, {Sizun}, {Pooley}, {Zdziarski},
  {Hannikainen}, {Kretschmar}, {Cordier}, \& {Lund}}]{Fuchs03}
{Fuchs}, Y., {Rodriguez}, J., {Mirabel}, I.~F., {et~al.} 2003, \aap, 409, L35

\bibitem[{{Gierli{\'n}ski} {et~al.}(1999){Gierli{\'n}ski}, {Zdziarski},
  {Poutanen}, {Coppi}, {Ebisawa}, \& {Johnson}}]{Gierlinski1999}
{Gierli{\'n}ski}, M., {Zdziarski}, A.~A., {Poutanen}, J., {et~al.} 1999,
  \mnras, 309, 496

\bibitem[{{Grinberg} {et~al.}(2011){Grinberg}, {Boeck}, {Pottschmidt},
  {Pooley}, {Wilms}, {Nowak}, {Cadolle Bel}, {Rodriguez}, {Marcu}, {Uttley},
  {Tomsick}, {Bodaghee}, \& {Markoff}}]{Grinberg2011}
{Grinberg}, V., {Boeck}, M., {Pottschmidt}, K., {et~al.} 2011, The Astronomer's
  Telegram, 3616, 1

\bibitem[{{Grinberg} {et~al.}(2013){Grinberg}, {Hell}, {Pottschmidt},
  {B{\"o}ck}, {Nowak}, {Rodriguez}, {Bodaghee}, {Cadolle Bel}, {Case}, {Hanke},
  {K{\"u}hnel}, {Markoff}, {Pooley}, {Rothschild}, {Tomsick}, {Wilson-Hodge},
  \& {Wilms}}]{Grinberg2013}
{Grinberg}, V., {Hell}, N., {Pottschmidt}, K., {et~al.} 2013, \aap, 554, A88

\bibitem[{{Grove} {et~al.}(1998){Grove}, {Johnson}, {Kroeger}, {McNaron-Brown},
  {Skibo}, \& {Phlips}}]{Grove1998}
{Grove}, J.~E., {Johnson}, W.~N., {Kroeger}, R.~A., {et~al.} 1998, \apj, 500,
  899

\bibitem[{{Houck} \& {Denicola}(2000)}]{Houck2000}
{Houck}, J.~C. \& {Denicola}, L.~A. 2000, in Astronomical Society of the
  Pacific Conference Series, Vol. 216, Astronomical Data Analysis Software and
  Systems IX, ed. N.~{Manset}, C.~{Veillet}, \& D.~{Crabtree}, 591

\bibitem[{{Jourdain} {et~al.}(2014){Jourdain}, {Roques}, \&
  {Chauvin}}]{Jourdain2014}
{Jourdain}, E., {Roques}, J.~P., \& {Chauvin}, M. 2014, \apj, 789, 26

\bibitem[{{Jourdain} {et~al.}(2012){Jourdain}, {Roques}, \&
  {Malzac}}]{Jourdain2012}
{Jourdain}, E., {Roques}, J.~P., \& {Malzac}, J. 2012, \apj, 744, 64

\bibitem[{{Kantzas} {et~al.}(2020){Kantzas}, {Markoff}, {Beuchert}, {Lucchini},
  {Chhotray}, {Ceccobello}, {Tetarenko}, {Miller-Jones}, {Bremer}, {Garcia},
  {Grinberg}, {Uttley}, \& {Wilms}}]{Kantzas2020}
{Kantzas}, D., {Markoff}, S., {Beuchert}, T., {et~al.} 2020, \mnras

\bibitem[{{Laurent} {et~al.}(2011){Laurent}, {Rodriguez}, {Wilms}, {Cadolle
  Bel}, {Pottschmidt}, \& {Grinberg}}]{Laurent2011}
{Laurent}, P., {Rodriguez}, J., {Wilms}, J., {et~al.} 2011, Science, 332, 438

\bibitem[{{Lebrun} {et~al.}(2003){Lebrun}, {Leray}, {Lavocat}, {Cr{\'e}tolle},
  {Arqu{\`e}s}, {Blondel}, {Bonnin}, {Bou{\`e}re}, {Cara}, {Chaleil}, {Daly},
  {Desages}, {Dzitko}, {Horeau}, {Laurent}, {Limousin}, {Mathy}, {Mauguen},
  {Meignier}, {Molini{\'e}}, {Poindron}, {Rouger}, {Sauvageon}, \&
  {Tourrette}}]{Lebrun2003}
{Lebrun}, F., {Leray}, J.~P., {Lavocat}, P., {et~al.} 2003, \aap, 411, L141

\bibitem[{{Lightman} \& {Zdziarski}(1987)}]{Lightman1987}
{Lightman}, A.~P. \& {Zdziarski}, A.~A. 1987, \apj, 319, 643

\bibitem[{{Lund} {et~al.}(2003){Lund}, {Budtz-J{\o}rgensen}, {Westergaard},
  {Brand t}, {Rasmussen}, {Hornstrup}, {Oxborrow}, {Chenevez}, {Jensen},
  {Laursen}, {Andersen}, {Mogensen}, {Rasmussen}, {Om{\o}}, {Pedersen},
  {Polny}, {Andersson}, {Andersson}, {K{\"a}m{\"a}r{\"a}inen}, {Vilhu},
  {Huovelin}, {Maisala}, {Morawski}, {Juchnikowski}, {Costa}, {Feroci},
  {Rubini}, {Rapisarda}, {Morelli}, {Carassiti}, {Frontera}, {Pelliciari},
  {Loffredo}, {Mart{\'\i}nez N{\'u}{\~n}ez}, {Reglero}, {Velasco}, {Larsson},
  {Svensson}, {Zdziarski}, {Castro-Tirado}, {Attina}, {Goria}, {Giulianelli},
  {Cordero}, {Rezazad}, {Schmidt}, {Carli}, {Gomez}, {Jensen}, {Sarri},
  {Tiemon}, {Orr}, {Much}, {Kretschmar}, \& {Schnopper}}]{Lund2003}
{Lund}, N., {Budtz-J{\o}rgensen}, C., {Westergaard}, N.~J., {et~al.} 2003,
  \aap, 411, L231

\bibitem[{{Malyshev} {et~al.}(2013){Malyshev}, {Zdziarski}, \&
  {Chernyakova}}]{Malyshev2013}
{Malyshev}, D., {Zdziarski}, A.~A., \& {Chernyakova}, M. 2013, \mnras, 434,
  2380

\bibitem[{{Malzac}(2013)}]{Malzac2013}
{Malzac}, J. 2013, \mnras, 429, L20

\bibitem[{{Malzac} \& {Belmont}(2009)}]{Malzac2009}
{Malzac}, J. \& {Belmont}, R. 2009, \mnras, 392, 570

\bibitem[{{McConnell} {et~al.}(2000){McConnell}, {Ryan}, {Collmar},
  {Sch{\"o}nfelder}, {Steinle}, {Strong}, {Bloemen}, {Hermsen}, {Kuiper},
  {Bennett}, {Phlips}, \& {Ling}}]{McConnell2000}
{McConnell}, M.~L., {Ryan}, J.~M., {Collmar}, W., {et~al.} 2000, \apj, 543, 928

\bibitem[{{McConnell} {et~al.}(2002){McConnell}, {Zdziarski}, {Bennett},
  {Bloemen}, {Collmar}, {Hermsen}, {Kuiper}, {Paciesas}, {Phlips}, {Poutanen},
  {Ryan}, {Sch{\"o}nfelder}, {Steinle}, \& {Strong}}]{McConnell2002}
{McConnell}, M.~L., {Zdziarski}, A.~A., {Bennett}, K., {et~al.} 2002, \apj,
  572, 984

\bibitem[{{Mirabel} {et~al.}(1992){Mirabel}, {Rodriguez}, {Cordier}, {Paul}, \&
  {Lebrun}}]{Mirabel92}
{Mirabel}, I.~F., {Rodriguez}, L.~F., {Cordier}, B., {Paul}, J., \& {Lebrun},
  F. 1992, \nat, 358, 215

\bibitem[{{Nowak} {et~al.}(2011){Nowak}, {Hanke}, {Trowbridge}, {Markoff},
  {Wilms}, {Pottschmidt}, {Coppi}, {Maitra}, {Davis}, \& {Tramper}}]{Nowak2011}
{Nowak}, M.~A., {Hanke}, M., {Trowbridge}, S.~N., {et~al.} 2011, \apj, 728, 13

\bibitem[{{Orosz} {et~al.}(2011){Orosz}, {McClintock}, {Aufdenberg},
  {Remillard}, {Reid}, {Narayan}, \& {Gou}}]{Orosz2011}
{Orosz}, J.~A., {McClintock}, J.~E., {Aufdenberg}, J.~P., {et~al.} 2011, \apj,
  742, 84

\bibitem[{{Poutanen} \& {Vurm}(2009)}]{Poutanen2009}
{Poutanen}, J. \& {Vurm}, I. 2009, \apjl, 690, L97

\bibitem[{{Reid} {et~al.}(2011){Reid}, {McClintock}, {Narayan}, {Gou},
  {Remillard}, \& {Orosz}}]{Reid2011}
{Reid}, M.~J., {McClintock}, J.~E., {Narayan}, R., {et~al.} 2011, \apj, 742, 83

\bibitem[{{Remillard} \& {McClintock}(2006)}]{Remillard2006}
{Remillard}, R.~A. \& {McClintock}, J.~E. 2006, \araa, 44, 49

\bibitem[{{Rodriguez} {et~al.}(2015){Rodriguez}, {Grinberg}, {Laurent},
  {Cadolle Bel}, {Pottschmidt}, {Pooley}, {Bodaghee}, {Wilms}, \&
  {Gouiff{\`e}s}}]{Rodriguez2015}
{Rodriguez}, J., {Grinberg}, V., {Laurent}, P., {et~al.} 2015, \apj, 807, 17

\bibitem[{{Romero} {et~al.}(2014){Romero}, {Vieyro}, \& {Chaty}}]{Romero2014}
{Romero}, G.~E., {Vieyro}, F.~L., \& {Chaty}, S. 2014, \aap, 562, L7

\bibitem[{{Roques} \& {Jourdain}(2019)}]{Roques2019}
{Roques}, J.-P. \& {Jourdain}, E. 2019, \apj, 870, 92

\bibitem[{{Rushton} {et~al.}(2012){Rushton}, {Miller-Jones}, {Campana},
  {Evangelista}, {Paragi}, {Maccarone}, {Pooley}, {Tudose}, {Fender},
  {Spencer}, \& {Dhawan}}]{Rushton2012}
{Rushton}, A., {Miller-Jones}, J.~C.~A., {Campana}, R., {et~al.} 2012, \mnras,
  419, 3194

\bibitem[{{Sabatini} {et~al.}(2013){Sabatini}, {Tavani}, {Coppi}, {Pooley},
  {Del Santo}, {Campana}, {Chen}, {Evangelista}, {Piano}, {Bulgarelli},
  {Cattaneo}, {Colafrancesco}, {Del Monte}, {Giuliani}, {Giusti}, {Longo},
  {Morselli}, {Pellizzoni}, {Pilia}, {Striani}, {Trifoglio}, \&
  {Vercellone}}]{Sabatini2013}
{Sabatini}, S., {Tavani}, M., {Coppi}, P., {et~al.} 2013, \apj, 766, 83

\bibitem[{{Siegert} {et~al.}(2019){Siegert}, {Diehl}, {Weinberger},
  {Pleintinger}, {Greiner}, \& {Zhang}}]{Siegert2019_SPIBG}
{Siegert}, T., {Diehl}, R., {Weinberger}, C., {et~al.} 2019, arXiv e-prints

\bibitem[{{Stirling} {et~al.}(2001){Stirling}, {Spencer}, {de la Force},
  {Garrett}, {Fender}, \& {Ogley}}]{Stirling2001}
{Stirling}, A.~M., {Spencer}, R.~E., {de la Force}, C.~J., {et~al.} 2001,
  \mnras, 327, 1273

\bibitem[{{Strong} {et~al.}(2005){Strong}, {Diehl}, {Halloin},
  {Sch{\"o}nfelder}, {Bouchet}, {Mandrou}, {Lebrun}, \&
  {Terrier}}]{Strong2005_gammaconti}
{Strong}, A.~W., {Diehl}, R., {Halloin}, H., {et~al.} 2005, \aap, 444, 495

\bibitem[{{Titarchuk}(1994)}]{Titarchuk1994}
{Titarchuk}, L. 1994, \apj, 434, 570

\bibitem[{{Vedrenne} {et~al.}(2003){Vedrenne}, {Roques}, {Sch{\"o}nfelder},
  {Mand rou}, {Lichti}, {von Kienlin}, {Cordier}, {Schanne}, {Kn{\"o}dlseder},
  {Skinner}, {Jean}, {Sanchez}, {Caraveo}, {Teegarden}, {von Ballmoos},
  {Bouchet}, {Paul}, {Matteson}, {Boggs}, {Wunderer}, {Leleux},
  {Weidenspointner}, {Durouchoux}, {Diehl}, {Strong}, {Cass{\'e}}, {Clair}, \&
  {Andr{\'e}}}]{Vedrenne2003}
{Vedrenne}, G., {Roques}, J.~P., {Sch{\"o}nfelder}, V., {et~al.} 2003, \aap,
  411, L63

\bibitem[{{Walter} \& {Xu}(2017)}]{Walter2017}
{Walter}, R. \& {Xu}, M. 2017, \aap, 603, A8

\bibitem[{{Wilms} {et~al.}(2006){Wilms}, {Nowak}, {Pottschmidt}, {Pooley}, \&
  {Fritz}}]{Wilms2006}
{Wilms}, J., {Nowak}, M.~A., {Pottschmidt}, K., {Pooley}, G.~G., \& {Fritz}, S.
  2006, \aap, 447, 245

\bibitem[{{Zanin} {et~al.}(2016){Zanin}, {Fern{\'a}ndez-Barral}, {de O{\~n}a
  Wilhelmi}, {Aharonian}, {Blanch}, {Bosch-Ramon}, \& {Galindo}}]{Zanin2016}
{Zanin}, R., {Fern{\'a}ndez-Barral}, A., {de O{\~n}a Wilhelmi}, E., {et~al.}
  2016, \aap, 596, A55

\bibitem[{{Zdziarski} {et~al.}(2012){Zdziarski}, {Lubi{\'n}ski}, \&
  {Sikora}}]{Zdziarski2012}
{Zdziarski}, A.~A., {Lubi{\'n}ski}, P., \& {Sikora}, M. 2012, \mnras, 423, 663

\bibitem[{{Zdziarski} {et~al.}(2017){Zdziarski}, {Malyshev}, {Chernyakova}, \&
  {Pooley}}]{Zdziarski2017}
{Zdziarski}, A.~A., {Malyshev}, D., {Chernyakova}, M., \& {Pooley}, G.~G. 2017,
  \mnras, 471, 3657

\bibitem[{{Zdziarski} {et~al.}(1998){Zdziarski}, {Poutanen}, {Mikolajewska},
  {Gierlinski}, {Ebisawa}, \& {Johnson}}]{Zdziarski1998}
{Zdziarski}, A.~A., {Poutanen}, J., {Mikolajewska}, J., {et~al.} 1998, \mnras,
  301, 435

\bibitem[{{Zdziarski} {et~al.}(2020){Zdziarski}, {Shapopi}, \&
  {Pooley}}]{Zdziarski2020}
{Zdziarski}, A.~A., {Shapopi}, J.~N.~S., \& {Pooley}, G.~G. 2020, \apjl, 894,
  L18

\end{thebibliography}

\appendix

\section{Variability timescales in Cyg X-1: the AIC map (SPI)}

\begin{figure*}[t]
\centering
\includegraphics[width=\textwidth]{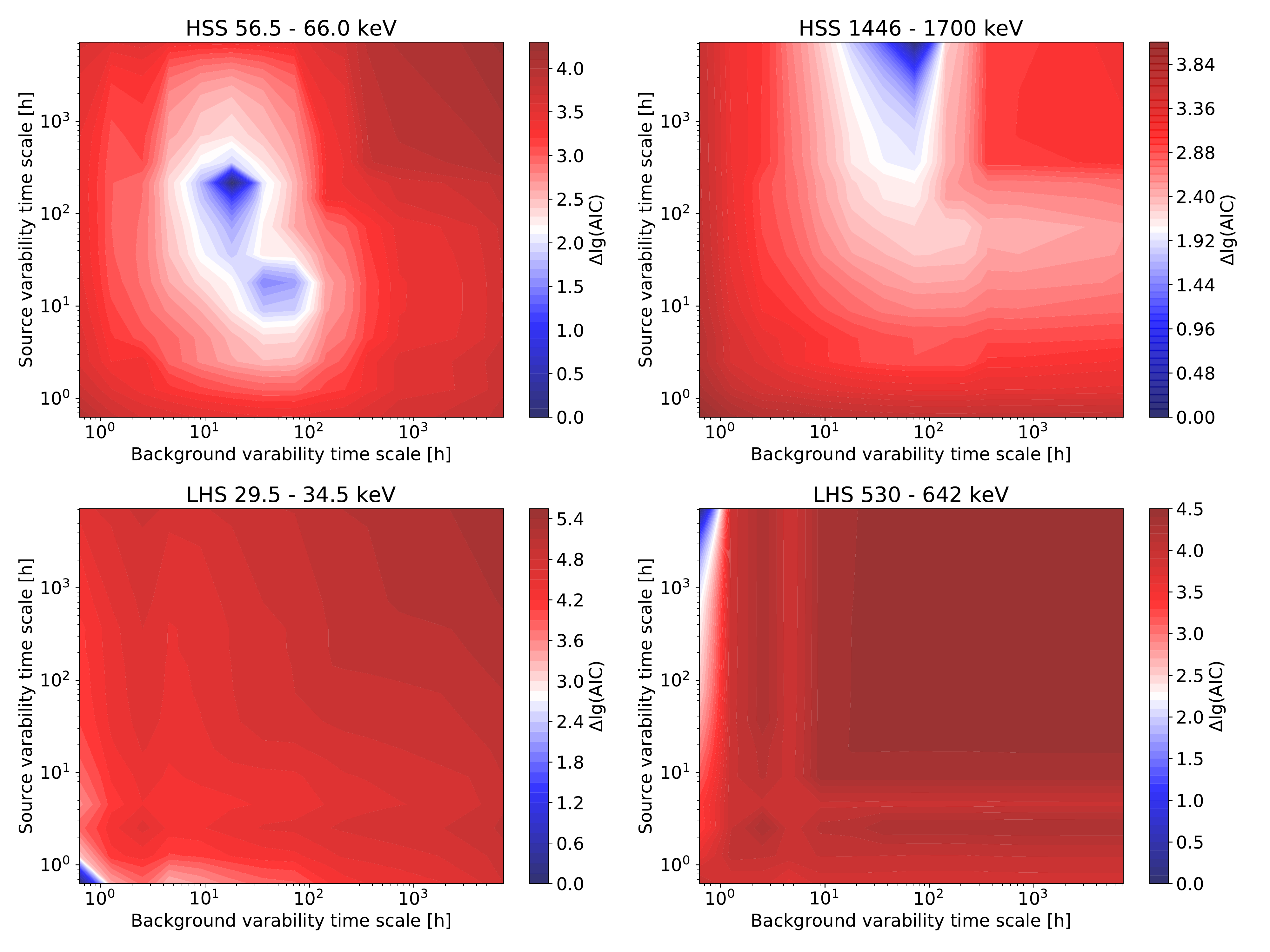}
\caption{Example of AIC maps with a different number of parameters for the background and the source variabilities.}
\label{afig:AIC_map}
\end{figure*}

Figure \ref{afig:AIC_map} shows AIC values as a result of maximum likelihood fits with different numbers of parameters for the background variability (timescale: the more parameters, the more variable) and the source (Cyg X-1) variability timescale, which was carried out in order to determine the flux and its uncertainty in each bin.
In the HSS, the AIC maps appear with a clear minimum (most probable solution according to AIC), slowly changing the source variability from $\sim$10\,h to constant as a function of energy.
In the LHS, the AIC minimum is found at the smallest timescale possible (one scw) for both source and background. For higher energies, the background still appears to vary on the shortest possible timescale, and the source seems constant. This appears unreasonable as the background variability should not change as a function of energy. We believe that the LHS spectral extraction suffers from a degeneracy in the determination of the number of required parameters. We therefore ``calibrate'' our spectrum using the IBIS spectral extraction for the LHS: using a source model without temporal variability but allowing the background to freely vary leads to an absolute normalization and shape close to the IBIS extraction.

\section{Comparison of different SPI spectral extraction methods}
\label{app:spi_extraction}

Here, we present two other spectral extraction methods for the SPI data. 

\subsection{Method 2: Using different event selection}
\label{sec:different_selection}

In this method, we consider the threshold of the PSD events, as set in the SPI telescope electronics. These are from 400\,keV to 2.2\,MeV after revolution 1170 and from nearly 650\,keV to 2.2\,MeV before. In this energy range, we used the PSD events only. This selection has been prescribed in \cite{Roques2019} in order to avoid spectral artifacts caused by potential spurious events. These spurious events appear at high energies ($>400$\,keV) for very bright sources (typically above 1 Crab). They are caused by a displacement of low-energy photons toward high energy, when the source is so bright that many photons can be detected in the same detection electronics time window. The energy reconstruction for the SE chain is then not reliable. This does not happen in the PE chain, which records the same event much more rapidly.





\subsection{Method 3: Using SPIDAI}

In this method, we use the SPI Data Analysis Interface\footnote{\url{https://sigma-2.cesr.fr/integral/spidai}.} (SPIDAI) to reduce the data from SPI. We created a sky model containing Cyg X-1 and Cygnus X-3 and set the source variability to ten scws for Cygnus X-3 and five scws for Cyg X-1 \citep[e.g.,][]{Bouchet2003, Cangemi2021}. We then created the background model by setting the variability timescale of the normalization of the background pattern to five scws. In order to avoid the effects due to solar flares, radiation belt entries, and other nonthermal incidents, we removed scws for which the reconstructed counts compared to the detector counts give a $\chi^2 > 1.5$. This selection reduces the total number of scws by $\sim 10$\,\%. The shadowgrams were then deconvolved to obtain the source flux, and spectra were extracted between 20 and 2000\,keV using 27 logaritmically spaced channels. 

Given the large number of scws in both states, we extracted six and four different spectra for the LHS and HSS, respectively, each of them was extracted from data of 500-600 scws and corresponds to a distinct period. For each spectrum, we applied a correcting factor of 1/0.85 \citep[above 400\,keV or 650\,keV according to the considered period,][]{Roques2019} to account for the change efficiency. We then computed two average spectra for both states. In the HSS, the normalization of the spectrum has a slight variability over the four different periods which explain the larger error bars that we observe in this state when computing the average spectrum.

\subsection{Comparison between the methods}
\begin{figure*}[t]
\centering
\includegraphics[width=\textwidth]{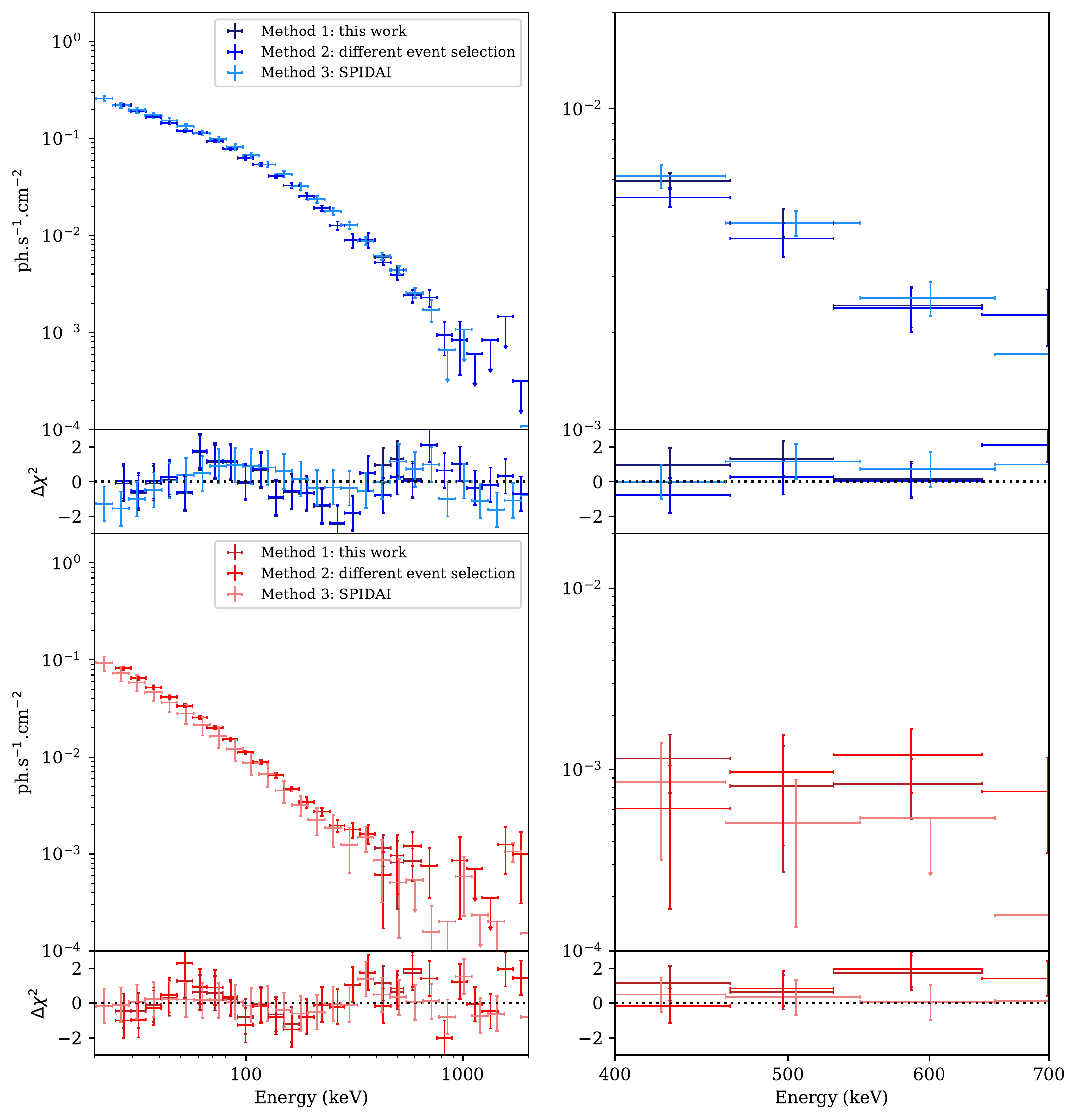}
\caption{\textit{Left pannel:} Spectra extracted with three different methods: method 1 described in Sec. \ref{sec:spi_data} of this paper (dark red), method 2 described in Sec. \ref{sec:different_selection} (medium red), and method 3 using SPIDAI (light red). \textit{Right pannel:} Focus on the 400-700\,keV energy range.}
\label{fig:comparison}
\end{figure*}

The right pannel on Fig. \ref{fig:comparison} shows the broad SPI spectra (20--2000\,keV) used in this paper along with the spectra extracted with the aforementioned methods. The left pannel on Fig. \ref{fig:comparison} shows the same spectra between 400\,keV and 700\,keV, that is to say the range for which the selection between SEs and PSD events is different from the the method described in this paper and method 2, described in Sec. \ref{sec:different_selection}.

In Fig. \ref{fig:comparison}, we show the residuals we obtain by fitting the spectra with a simple cutoff powerlaw (\texttt{constant*cutoffpl}, $\chi^2_{\mathrm{LHS}} = 75.50/76$, $\chi^2_{\mathrm{HSS}} = 93.65/76$). By fixing the constant parameter of the spectra used in this paper to 1, we obtain constant values of $c_{\mathrm{method\ 2}} = 0.97 \pm 0.02$ and $c_{\mathrm{method\ 3}} = 1.06 \pm 0.03$ for the LHS and $c_{\mathrm{method\ 2}} = 1.00 \pm 0.02$ and $c_{\mathrm{method\ 3}} = 0.86 \pm 0.06$ for the HSS. For both states, the three spectra are consistent and we do not observe any significant differences, in particular in the 400--650 keV range, the range most affected by spectral artifacts caused by potential spurious events.

We also investigate whether the parameters extracted in this study would be different using method 2 or method 3 spectrum. We find that the parameters remain unchanged and zre consistent within the uncertainties for both states when applying the semi-physical model or the hybrid corona models (\texttt{eqpair} and \texttt{belm}).

\end{document}